\documentclass[12pt]{article}
\usepackage[hmargin=2.7cm,vmargin=2.3cm]{geometry}
 \usepackage{graphicx}
\usepackage{amsmath,amsfonts,amsthm}
 \usepackage{amssymb}
  \usepackage{enumerate}
\usepackage{cite}
\usepackage{url}
\newcommand{\bey}{\begin{eqnarray}}
\newcommand{\eey}{\end{eqnarray}}
\usepackage{etoolbox}
\usepackage{subfig}
\usepackage{lipsum}

\newcommand{\bfnu}{\boldsymbol{\nu}}
\newcommand{\R}{\mathbb{R}}
\newcommand{\C}{\mathbb{C}}

\begin{document}
\title{Final States in Quantum Cosmology: Cosmic Acceleration as a Quantum Post-Selection Effect}
\author{ Charis Anastopoulos\footnote{anastop@upatras.gr} \\
 Department of Physics, University of Patras, 26500 Greece} 
\maketitle

\begin{abstract}
Standard quantum theory admits naturally statistical ensembles that are both pre-selected and post-selected, i.e., they involve both an initial and a final state. We argue that there is no compelling physical reason to preclude a probability assignment with a final
quantum state at the cosmological level. We therefore analyze the implications of a final state in the probability assignment for quantum cosmology. To this end, we derive effective classical equations of motion for systems subject to both initial and final conditions. Remarkably, these effective equations do not depend on the details of the quantum theory, but only on the geometric features of the classical state space. When applied to  Friedman-Robertson-Walker cosmological models, these effective equations generically describe cosmic acceleration in the absence of a cosmological constant, dark energy, or modified gravitational dynamics. Therefore, cosmic acceleration emerges as a  quantum post-selection effect, that is, a macroscopic quantum phenomenon.
\end{abstract}

\section{Introduction}
Quantum measurements are usually described in terms of {\em pre-selected} ensembles, that is, quantum ensembles whose elements have been selected prior to the measurement. Information about pre-selection is encoded in the quantum state. Like all probabilistic theories, quantum theory also admits {\em post-selected} ensembles \cite{ABL, twostate}, that is, quantum ensembles in which we discard all measurement outcomes unless a final condition is satisfied. 

Quantum theory treats pre-selection and post-selection symmetrically. Both are expressed in terms of  density matrices: an initial density matrix  $\hat{\rho}_0$ for pre-selection at time $t = 0$, and a final density matrix $\hat{\rho}_f$ for post-selection at time $ T$. Suppose that at time $t \in [0, T]$, we measure an observable that corresponds to a self-adjoint operator $\hat{A} = \sum_i a_i \hat{P}_i$, where  $\hat{P}_i$ are the associated spectral projectors. The probability that the measurement gives value $\hat{a}_i$ is given by
\bey
\mbox{Prob}(a_i, t) = C \mbox{Tr} \left[ e^{-i\hat{H}(T - t)}\hat{P}_ie^{-i\hat{H}t} \hat{\rho}_0e^{i\hat{H}t} \hat{P}_ie^{i\hat{H}(T - t)} \hat{\rho}_f\right],\label{prepost}
\eey
where $\hat{H}$ is the Hamiltonian of the system, and $C > 0$ a normalization constant. Note that the probabilities are invariant under the exchange $t \leftrightarrow T-t$ and $\hat{\rho}_0 \leftrightarrow \hat{\rho}_f$.

The probabilities (\ref{prepost}) have a clear operational meaning when applied to measurements in a laboratory, and they have been employed in a variety of contexts---see, for example, Refs. \cite{AAV, AV90, Knill, Aaronson, SGV, TC13}. In this paper, we will analyze their implications for cosmology. Quantum cosmology aims to describe the Universe as a closed quantum system, by assuming that the usual rules of quantum theory can be upgraded to work beyond a strictly operational framework. In the quantum cosmological setting, one usually employs the quantum probability rule for pre-selected ensembles. This means that absolute probabilities---that is, probabilities prior to any conditioning---are defined with reference to an initial quantum state of the Universe and no final state. Post-selection is only applied to conditional probabilities, for example, in the context of anthropic arguments \cite{anthropic}.

In classical mechanics, we can either postulate initial conditions for all  degrees of freedom of a physical system (as in Hamiltonian mechanics), or subject some degrees of freedom  to initial conditions and others to final conditions (as in the least action principle). This difference is a matter of convention or practicality \cite{YoMa}, and it does not affect physical predictions.  
However, there is a fundamental intuition of a logical/causal arrow of time, namely, that the cause precedes the effect. This implies that 
the present state of the Universe follows of an earlier state, and this from the evolution of a yet earlier state, back to the beginning of time. This is strong reason to believe that in cosmology, the initial value problem provides a more fundamental description.

The intuition of the causal arrow of time, as described above, refers explicitly to the evolution of classical  microstates, which are represented by   points in a classical state space. It does not work well with quantum states \cite{Isham}, which are radically different. The key difference is that any two classical microstates are mutually exclusive, but two non-orthogonal quantum states are not; this prevents us from viewing the quantum state as an objective feature of an individual quantum system \cite{Ana23}.
 Furthermore, the violation of local realism by Bell's inequalities requires a reexamination of many notions of classical causation \cite{healey}. In most interpretations of quantum theory---including the Copenhagen interpretation---the quantum state is viewed solely as an informational object that determines probabilities. In the laboratory, a choice of initial conditions reflects the preparation of the system by the experimentalist, i.e., it reflects operations by agents external to the system. In quantum cosmology, state preparation is not an option, so we should consider all possible rules of probability assignment permissible by the quantum formalism.
For this reason, the idea that only initial conditions are meaningful loses much of its force  in quantum cosmology.  In histories-based formulations of quantum theory \cite{Gri, Omn1, Omn2, GeHa1, hartlelo, Ish94, Sork, Sav10}, the quantum state is not a fundamental object, and it is possible to define probabilities that involve both initial and  final conditions, and also intermediate ones. Each choice of conditions  defines a different  
cosmological theory that must be judged on its own merits, that is, logical consistency and agreement with observation.

Final states in a cosmological setting have been considered before. Following the pioneering paper of Aharonov, Bergmann and Lebowitz \cite{ABL},  Cocke considered initial and final conditions in oscillating cosmological models \cite{cocke}.   Schulman  also analyzed the effect of future conditions on current observables, mainly with an emphasis on the arrow of time \cite{Schul1, Schul2, Schul3, Schul4}.

In  quantum cosmology, the discussion of final states grew out of
 Hartle and Hawking's no-boundary proposal \cite{HH83, HH85} about the wave function of the Universe. The no-boundary proposal involves a path integral with boundary conditions that, in principle, apply to both early and late times \cite{Haw85, Pa85, HLL}. This feature motivated the later consideration of ``time-neutral cosmologies" by Gell-Mann and Hartle \cite{GeHa94, Har20}, in which $\hat{\rho}_f \neq \hat{I}$ in Eq. (\ref{prepost})---see also Refs. \cite{Page93, Kent, craig}. Note that a common motif in the above works is the assumption of a closed universe, as the final conditions were imposed at the final singularity. Particular emphasis was given on the possibility that the thermodynamic arrow of time could coincides with the cosmological arrow of time (entropy decrease in the collapsing phase of the Universe) \cite{Gold}. The discovery of cosmic acceleration---implying an open universe---undermined to some extent the theoretical motivation for this line of research.
  
Following a different rationale,  Davies studied the consequences of a final condition in relation to the cosmological particle creation, and he identified effects that could be observable at the present cosmological epoch \cite{Dav13}. 
We also note that cosmological  calculations invoking the anthropic principle \cite{Carter, CaRe, Barrow, anthropic} may involve probabilities conditioned at later times. In a non-cosmological context,  the imposition of a final condition at black hole singularities has been proposed as a way to reconcile information loss with string theory \cite{HoMa}.

Here, we focus on a novel aspect of quantum systems with both pre- and post-selection: their classical limit may  strongly diverge from the usual classical equations of motion. In particular, we derive effective quasiclassical equations of motion \cite{GeHa2} for a large class of systems through the analysis of post-selected probabilities associated to coarse-grained observables. We show the existence of a {\em hidden determinism} among the rare quantum processes that are probed by post-selection. 
The effective equations that we derive do not depend on the detailed properties of the quantum description. For a large class of systems, they only depend on the geometric features of the classical state space.
 When we apply this formalism to simple quantum cosmological models, we find that post-selection in
Friedmann-Robertson-Walker (FRW) cosmology naturally leads to cosmic acceleration in absence of a cosmological constant, dark energy, or modified gravity. 
Hence,  cosmic acceleration   emerges as a quantum ``post-selection" effect.

The structure of this paper is the following. In Sec. 2, we derive the general equations of motion for a quantum system that is both pre- and post-selected. To this end, we assume that the initial and final condition is well localized in state space, and that the quantum dynamics reduce to classical Hamiltonian dynamics for sufficiently coarse-grained observables. Remarkably, the effective equations carry little to no memory of the quantum theory, and can only be expressed in terms of geometric properties of the classical state space. 
In Sec. 3, we briefly present the decoherent histories approach, which provides the most general framework for the discussion of final conditions in quantum cosmology.
The material in this section is mostly of a review character, but they are needed in order to provide the background to the following section, which deals with 
 quantum cosmology.  In Sec. 4, we show that the effective equations of motion with post-selection in FRW cosmology correspond to  cosmic acceleration. In Sec. 5, we discuss our results and their implications.

\section{The classical limit in a post-selected system}

In this section, we derive effective classical equations of motion for post-selected quantum systems. We find that under very mild assumptions, these effective equations do not depend on the details of the quantum theory, and they can be expressed solely in terms of geometric quantities on the classical state space.

\subsection{The classical limit of a quantum system}

There are several different methods for obtaining the classical limit of a quantum system. The simplest ones involve monitoring the evolution of the expectation value of selected variables (as in Ehrenfest's theorem), or the time evolution of peaks in the wave function. More elaborate methods involve the construction of quantum observables that evolve almost deterministically (i.e., they are correlated by classical equations of motion). The most general method is the construction of probabilities for histories, and the identification of the histories that maximize them.

Not all methods  work in presence of post-selection. For example, there is no Ehrenfest's theorem. To see this, consider 
an observable $\hat{A} = \sum_i a_i \hat{P}_i$, where $a_i$ are its eigenvalues and $\hat{P}_i$ are its spectral projectors. The expectation value of $\hat{A}$ at time $t$ is $\langle \hat{A}(t) \rangle = \sum_i a_i \mbox{Prob}(a_i, t)$, where $\mbox{Prob}(a_i, t)$ are the probabilities (\ref{prepost}). Since $\mbox{Prob}(a_i, t)$ is not linear with respect to $\hat{P}_i$, the expectation value of $\hat{A}$ is not a linear functional of $\hat{A}$. Hence, the derivation of Ehrenfest's theorem does not pass through.

The simplest method that works in  post-selected systems is finding how the  peaks in the probability distribution (\ref{prepost}) depend on the time parameter $t$.  In the Appendix A, we  present two simple examples that use this method: a classical stochastic system and a quantum particle. The method is simple to implement, but its scope is limited. The main problem is that in most systems of relevance, time evolution leads to a probability distribution with multiple peaks or a  large spread, so that the behavior of a single peak has  little relevant information. Furthermore, as shown in the example of Chapter \ref{peaks}, there is no guarantee that the classical paths determined by the peaks will satisfy the boundary conditions. 

A mathematically and conceptually rigorous derivation of the classical limit requires a consistent implementation of {\em coarse-graining} in the quantum system. This requires the definition of macroscopic quantum observables, that is, quantum observables that describe macroscopic properties of the system. These observables are stable under  Heisenberg-time  evolution, and they are correlated in time according to a classical, approximately deterministic evolution law. This method enables us to derive rigorously Hamiltonian dynamics as a limit of quantum unitary evolution.  For a detailed presentation, see Chapter 6 of Ref. \cite{Omn1} and Ref. \cite{Omnes89}; for an extension to open quantum systems and the derivation of effective classical dynamics with dissipation, see Ref. \cite{Ana96}; 
  for an analogous construction in classical statistical mechanics, see  Ref. \cite{OPen}.

\subsection{ Coarse-grained observables}

Consider a quantum system described by a Hilbert space ${\cal H}$. We define coarse-grained observables   in terms of a Positive-Operator-Valued measures (POVMs) on ${\cal H}$. A POVM is a family of positive operators $\hat{\Pi}_{\xi}$, where $\xi$ takes values in a set   $\Gamma$ of classical alternatives.  The POVM is normalized to unity, that is, 
$\int d\xi \hat{\Pi}_{\xi} = \hat{I}$; here, $d \xi $ is an appropriate integration measure on $\Gamma$.

To define approximate deterministic behavior, we consider 
  large subsets $V$ of $\Gamma$, such that the operators $\hat{\Pi}_V = \int_V d\xi \hat{\Pi}_{\xi}$ are {\em approximate projectors}. This means  that  $\hat{\Pi}_V$ satisfies 
\bey
Tr|\hat{\Pi}_V - \hat{\Pi}_V^2| < \epsilon Tr \hat{\Pi}_V. \label{approj}
\eey
for some $\epsilon << 1$. Physically, this condition implies that  second measurement of $V$ immediately after the first will give the same result with accuracy of order $\epsilon$. We call  subsets $V$ of $\Gamma$ that satisfy Eq. (\ref{approj}) $\epsilon$-regular or simply regular.

Let $\hat{H}$ be the Hamiltonian of the system, and $\hat{\Pi}_V(t) = e^{i\hat{H}t}\hat{\Pi}_Ve^{-i\hat{H}t}$ be the Heisenberg-picture evolution of  $\hat{\Pi}_V$. A {\em coarse-grained} observable is a exclusive and exhaustive family of regular subsets $V_a$ of $\Gamma$, that is, a family of regular subsets, such that 
 $\cup_aV_a = \Gamma$, and   $V_a \cap V_{a'} = \emptyset$ for $a \neq a'$. A coarse-grained observable $\{V_a\}$ 
exhibits approximate determinism in the time interval $[0, T]$, if there exists  a family of functions $f_t:\Gamma \rightarrow \Gamma$ for $t \in [0, T]$, such that 
\bey
\hat{\Pi}_{V_a}(t) \simeq \hat{\Pi}_{f_{-t}V_a} \label{approxeq}
\eey
 for all $a$ and for all $t$. This means that Heisenberg-picture time evolution induces a deterministic law $f_t$ of time evolution for the alternatives $V_a$ defined on the space $\Gamma$ of classical alternatives\footnote{We use   $f_{-t}$ in Eq. (\ref{approxeq}) because the relevant geometric quantity is the pull-back $f^*_t\chi_V$ of the characteristic function of $V$, defined as $f^*_t\chi_V(\xi) = \chi_V(f_t(\xi))$. }.

In Eq. (\ref{approxeq}), we used the notion of approximate equality heuristically. Eq. (\ref{approxeq}) can be expressed rigorously with reference to an appropriate Hilbert space norm. For example, using the trace norm, we express Eq. (\ref{approxeq}) as
\bey
Tr|\hat{\Pi}_{V_a}(t) - \hat{\Pi}_{f_{-t}(V_a)}| < \epsilon Tr \hat{\Pi}_{V_a}, \label{approxdet}
\eey
for some $\epsilon << 1$.

\subsection{Coherent states} 

 We will analyze  approximate determinism in the classical state space $\Gamma$. We will construct a  POVM for state space observables through the 
introduction of (generalized) coherent states $|\xi\rangle $ on ${\cal H}$. A set of coherent states  is a continuous family of vectors $|\xi \rangle \in {\cal H}$, labeled by the points $\xi \in \Gamma$, which satisfy the  over-completeness relation
 $\int d \xi |\xi \rangle \langle \xi| = \hat{I}$ for some integration measure $d\xi$ on $\Gamma$ \cite{Klauder}.
 
 Typically, generalized coherent states are defined in terms of group representations, i.e., each $|\xi\rangle$ can be written as $\hat{U}(g)|0\rangle$, where $\hat{U}(g)$ is a unitary irreducible representation of a group $G$ and $|0\rangle$ is a reference vector \cite{Perelomov}. We define the subgroup $H $
  of $G$ as   $H = \{ h\in G; |\langle 0|\hat{U}(h)|0\rangle| = 1\}$. Then $\Gamma$ is the quotient manifold $G/H$, and it is a homogeneous space for the group $G$.

  Coherent states also induce a Riemannian metric 
\bey
\gamma_{ab} = - \langle \xi|\partial_a\partial_b|\xi\rangle + \langle \xi| \partial_a|\xi\rangle \langle \xi | \partial_b| \xi\rangle, \label{metric}
\eey  
and a two-form $\omega = -i d \langle \xi|d|\xi\rangle$ on $\Gamma$.  The most important class of coherent  states is characterized by non-degenerate $\omega$. Then, $\Gamma$ is a homogeneous symplectic manifold. The irreducible representation of $G$ on the Hilbert space defines an action of $G$ on $\Gamma$ that preserves both the symplectic form (hence, it corresponds to canonical transformations) and the metric (hence, it defines a group of isometries). 
 The integration measure $d \xi$ coincides with the standard Liouville measure  of a symplectic manifold. In this case, $\Gamma$ can be interpreted as the state space of a classical dynamical system with $G$ as a symmetry group \cite{MaRa}.
The metric (\ref{metric}) 
defines a distance function on $\Gamma$, and it enables us to define dimensionless coordinates $\xi^a$ in units where $\hbar = 1$.  Here, we will 
 denote the distance between two points $\xi_1$ and $\xi_2$ by $||\xi_1 - \xi_2||$.
  
  For a non-compact $\Gamma$, the inner product $\langle \xi_1|\xi_2\rangle$ is strongly suppressed with the distance $||\xi_1 - \xi_2||$. This means that 
 \bey
 |\langle \xi_1|\xi_2\rangle|^2 \leq \upsilon\left(||\xi_1 - \xi_2||\right), \label{inprod}
\eey
where $\upsilon(x)$ is a function that vanishes fast for $x >> 1$.
  
 The canonical coherent states are defined on the Hilbert space $L^2({\bf R}^n)$, in which case $\xi \in {\bf C}^n$. They satisfy  
 \bey
 \langle \xi_1|\xi_2\rangle = \exp\left[ \frac{1}{2}(\xi_1^* \cdot \xi_2 - \xi_2^* \cdot \xi_1) - \frac{1}{2} |\xi_1 - \xi_2|^2\right]
 \eey
 where $\cdot$ stands for the Euclidean inner product on ${\bf C}^n$ and $|\xi| = \sqrt{\xi^* \cdot \xi}$ is the associated norm. Hence, we have equality in Eq. (\ref{inprod}), with $\upsilon(x) = e^{-x^2}$.

\subsection{Deriving approximate determinism} 
Given a set of coherent states $|\xi\rangle$ on a Hilbert space ${\cal H}$, we define the coarse-grained observables $\hat{\Pi}_{\xi} = |\xi\rangle \langle \xi|$. Then, the observable associated to a region $V \subset \Gamma$, $\hat{\Pi}_V = \int_V d\xi |\xi\rangle \langle \xi|$ satisfies $Tr \hat{\Pi}_V = [V]$, where $[V] = \int_V d \xi$ is the volume of $V$.

\begin{figure}[!tbp]
 \includegraphics[width=0.9\textwidth]{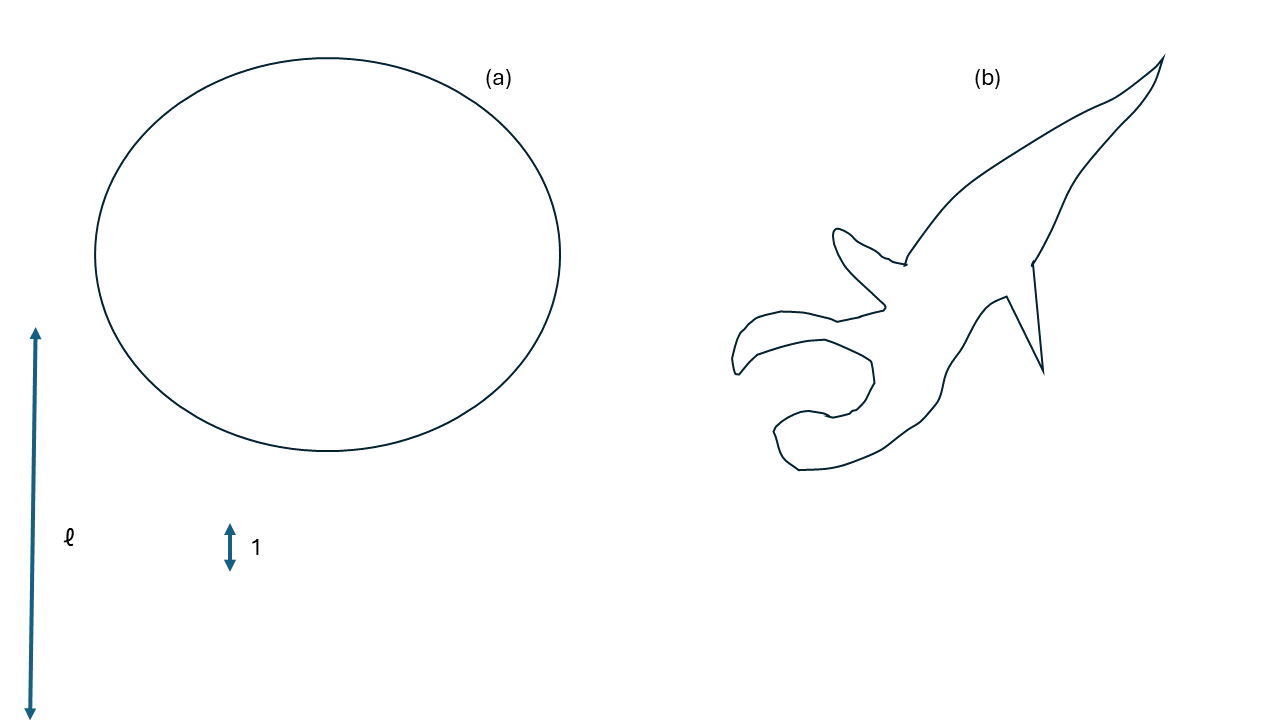}
  \caption{ 
   (a) A regular cell with typical size larger than $\ell >> 1$ in all directions. (b) A non-regular cell with some characteristic scales of the order of unity. We use dimensionless units with $\hbar = 1$. If $L$ is the typical length scale and $P$ the typical momentum scale, then $\ell  = \sqrt{LP/\hbar}$. 
    }
      \label{regc}
\end{figure}

Consider a subset  $V$ of $\Gamma$ with typical dimensions are larger than $\ell >> 1$---see Fig. \ref{regc}.  
 The  volume of $V$ is $[V] = c_1 \ell^{\nu}$, where $c_1$ is a constant, $\nu = \mbox{dim}\Gamma$, and $c_1$ is a number of order unity. 
Such subsets are regular. The operators $\hat{\Pi}_V$ are approximate projectors: they satisfy Eq. (\ref{approj}) with $\epsilon \sim \ell^{-1}$.
  
To see this, note that $\hat{\Pi}_V - \hat{\Pi}_V^2 = \hat{\Pi}_V \hat{\Pi}_{\Gamma - V} = 
\int_V d\xi \int_{\Gamma - V} d \xi' \langle \xi|\xi'\rangle |\xi\rangle \langle \xi'|$. All terms with $||\xi - \xi'|| >> 1$ are suppressed with $\upsilon\left(||\xi - \xi'|| \right)$. The only terms for which $|\xi - \xi'|$ is of order unity or smaller come from a strip $M$ around the boundary $\partial V$. The width of the strip is of order unity, hence, $[M]$ is of the same order as the area of $\partial V$. But the latter is at most equal to $[V]/\ell$. Hence, $[M]/[V] = c'\ell^{-1}$, where $c'$ is of order unity. The dominant contribution to $Tr|\hat{\Pi}_V - \hat{\Pi}_V^2|/Tr\hat{\Pi}_V$ is proportional to $[M]/[V]$, hence to $\ell^{-1}$.

Similarly, we can show that for two disjoint regular regions $V$ and $V'$ of typical size $\ell$, 
\bey
\frac{Tr|\hat{\Pi}_V - \hat{\Pi}_{V'}|}{Tr \hat{\Pi}_V} < c_2\ell^{-1}, 
\eey
where $c_2$ is a constant of order unity. 
Hence, operators corresponding to mutually exclusive subsets of $\Gamma$ are approximately disjoint. This guarantees that we can interpret an operator 
$\hat{\Pi}_V$ as representing the  statement that the system is found in the subset $V$ of $\Gamma$ up to an error of order $\ell^{-1}$. For rigorous proofs of these statements, see Ref. \cite{Omnes89}.

For the special case of canonical coherent states, and for a Hamiltonian $\hat{H}$ of the form $\sum_a \frac{\hat{p}_a^2}{2m_a} + V(\hat{x}_a)$, Omn\'es showed that coarse-grained observables defined by regular sets $V_a$ satisfy approximate determinism \cite{Omnes89}. That is, the associated approximate projectors $\hat{\Pi}_{V_a}$ satisfy Eq. (\ref{approxdet}), where $f_t$ stands for the equations of motion with respect to the classical Hamiltonian $h = \sum_a \frac{p_a^2}{2m_a} + V(x)$. Approximate determinism holds for all $T$ such that the set $f_{-t}(V)$ remains regular.

We use this result, in order to identify an almost deterministic evolution law for post-selected evolution of state space observables. To this end, we partition the time interval $[0, T]$ into $N$ steps, and we consider $N - 1$ successive   measurements at times $t_n = n T/N$, $n = 1, 2, \ldots, N-1$, which correspond to  
 state space regions $V_1, V_2, \ldots V_{N-1}$. Given an initial state $|\psi_i\rangle$ and a final state $|\psi_f\rangle$, the  probability $\mbox{Prob}(V_1,t_1; V_2,t_2; \ldots; V_{N-1}, t_{N-1})$ associated to those measurements is 
\bey
\mbox{Prob}(V_1,t_1; V_2,t_2; \ldots; V_{N-1}, t_{N-1}) = K |{\cal A}(V_1,t_1; V_2,t_2; \ldots; V_{N-1}, t_{N-1})|^2, \label{probdis}
\eey
where $K$ is a normalization constant, and 
\bey
{\cal A}(V_1,t_1; V_2,t_2; \ldots; V_{N-1}, t_{N-1}) = 
\langle \psi_f|e^{i\hat{H}T} \hat{\Pi}_{V_{N-1}}(t_{N-1}) \ldots \hat{\Pi}_{V_{2}}(t_{2}) \hat{\Pi}_{V_{1}}(t_{1})|\psi_i\rangle \nonumber \\
\simeq 
\langle \psi_f'| \hat{\Pi}_{f_{-t_{N-1}}(V_{N-1})} \ldots \hat{\Pi}_{f_{-t_2}(V_2)}  \hat{\Pi}_{f_{-t_1}(V_1)}|\psi_i\rangle,
\eey
is a probability amplitude. Here,  
 we wrote $|\psi'_f \rangle = e^{i\hat{H}T} |\psi_f\rangle$. We assume that $|\psi_i\rangle$ and $|\psi_f'\rangle$ are well localized around state space points $\xi_i$ and $\xi_f'$, respectively. The classical limit is not affected if we take those states to be coherent states, $|\psi_i \rangle = |\xi_i\rangle$ and $|\psi_f'\rangle
= |\xi'_f\rangle$.  Note that by Ehrenfest's theorem, $\xi'_f = f_{-T}(\xi_f)$, where $\xi_f$ is the localization center of $|\psi_f\rangle$.

We write $ \xi_0 = \xi_i $ and $= \xi_N = \xi_f'$, so that
\bey
{\cal A}(V_1,t_1; V_2,t_2; \ldots; V_{N-1}, t_{N-1}) = \int_{f_{-t_1}(V_1)} d\xi_1 \int_{f_{-t_2}(V_2)} d\xi_2 \ldots  \int_{f_{-t_{N-1}}(V_{N-1})} d\xi_{N-1} \prod_{n=0}^{N-1}\langle \xi_n|\xi_{n+1}\rangle \nonumber \\
= \int_{f_{-t_1}(V_1) \times f_{-t_2}(V_2) \times \ldots \times f_{-t_{N-1}}} d^{N-1}\xi  
\exp\left(\frac{i}{2} \sum_{n=0}^{N-1} \mbox{Im}({\bf \xi}_n^*\cdot {\bf \xi}_{n+1} - \frac{1}{2} \sum_{n=0}^{N-1} |\xi_n - \xi_{n+1}|^2 \right).\hspace{1.4cm} \label{AVN}
\eey
Consider now a partition of $\Gamma$ into mutually exclusive and exhaustive cells $V_{\alpha}$ of typical size $\ell >> 1$, where $\alpha$ runs in a set of indices $L$. 
Eq. (\ref{probdis}) defines a probability density on the set $H$ of histories, that is, of sequences of cells $(V_{\alpha_1}, V_{\alpha_2}, \ldots, V_{\alpha_{N-1}})$ associated to the times $t_1, t_2, \ldots, t_{N-1}$. The key point here is that the probability distribution is peaked around a single history containing the unique discrete path $\xi_n = \bar{\xi}_n$ that minimizes the term $\sum_{n=0}^{N-1} |\xi_n - \xi_{n+1}|^2$ in Eq. (\ref{AVN}). Any other history involves integration over paths with at least one point at distance greater than $\ell$ from $\xi_n$, hence, the associated probability is suppressed by a term of order $e^{-\ell^2}$ or worse. We conclude that the probability measure (\ref{probdis}) is characterized by approximate determinism: it assigns probability very close to one to a single element.

\begin{figure}[!tbp]
 \includegraphics[width=1\textwidth]{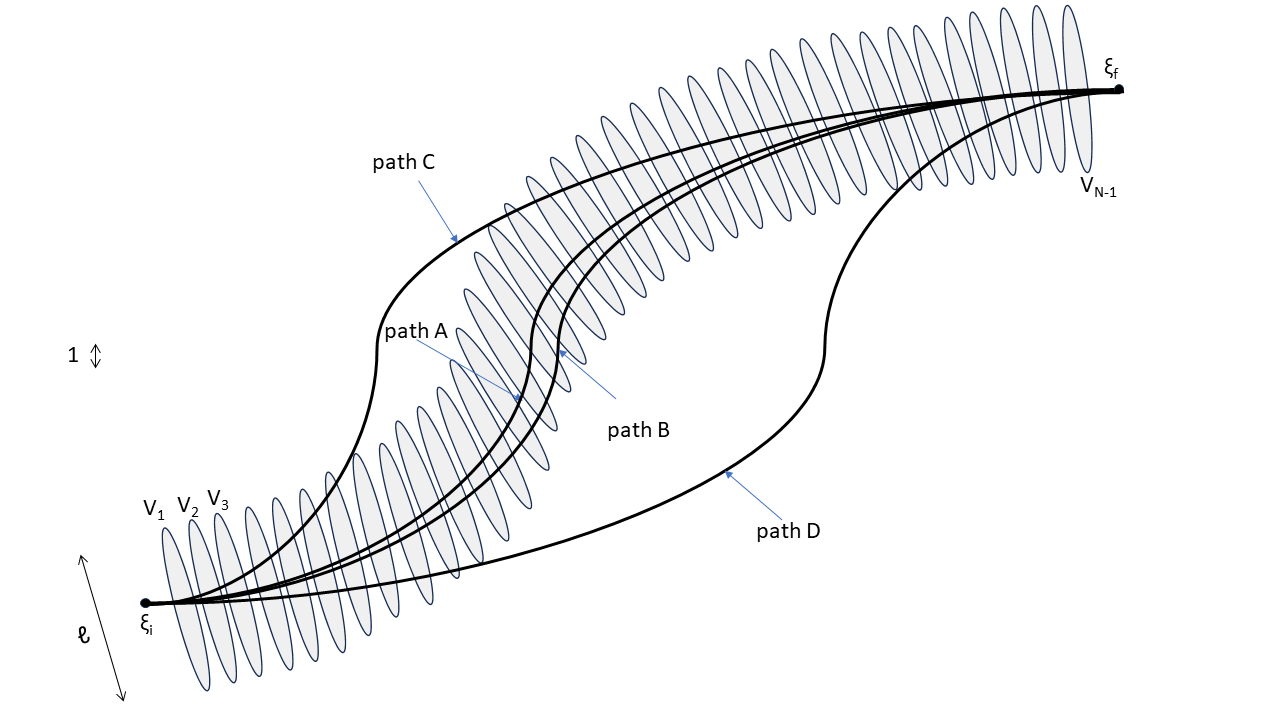}
  \caption{ 
   A  sequence of coarse-grained state space measurements corresponds to a tube around a classical path. The typical scale $\ell >> 1$ of the cells is much larger than the length-scale of the underlying metric. The probabilities is close to unity that the history $V_1, V_2, \ldots, V_{N-1}$ is realized. Path A represents the path (\ref{traj}); it is classically indistinguishable from path B that also lies within the same coarse-grained history.  Paths C and D contribute to different coarse-grained histories, say $W_1, W_2, \ldots, W_{N-1}$ with probabilities suppressed by a factor of at least
      $e^{-\ell^2}$.}
      \label{figaa}
\end{figure}

To identify this element, we minimize $\sum_{n=0}^{N-1} |\xi_n - \xi_{n+1}|^2$ with respect to $\xi_n, n = 1, 2, \ldots N-1$. We obtain the difference equation $\bar{\xi}_n = \frac{1}{2} (\bar{\xi}_{n-1} + \bar{\xi}_{n+1})$, with solution 
\bey
\bar{\xi}_n = \xi_i + (\xi_f' - \xi_i) n/N. \label{xin}
\eey
 This path is contained in the sequence of cells $(f_{-t_1}(V_1), f_{-t_2}(V_2), \ldots, f_{-t_{N-1}}(V_{N-1}))$. The corresponding path with respect to the sequence of cells $V_1, V_2, \ldots, V_{N-1}$ is 
\bey
\xi_n = f_{t_n}(\bar{\xi}_n) = f_{t_n}[\xi_i + (\xi_f' - \xi_i) n/N] = f_{t_n}\left[\xi_i + (f_{-T}(\xi_f) - \xi_i) \frac{n}{N}\right].
\eey
At the continuous limit,  we obtain
\bey
\xi(t) = f_t\left(\xi_i + (f_{-T}(\xi_f) - \xi_i)t/T\right). \label{traj}
\eey 
Eq. (\ref{traj}) is the main result of this section. It provides a unique classical trajectory derived from the post-selected quantum system under coarse-graining. Strictly speaking, Eq. (\ref{traj}) defines a phase space tube of width $\ell$, as seen in Fig. \ref{figaa}. All paths that lie within this tube are macroscopically indistinguishable from the path (\ref{traj}).

An important feature of Eq. (\ref{traj}) is that it is expressed solely in terms of objects on the classical state space. This is remarkable, because the notion of 
post-selected time evolution makes no sense in a deterministic theory. We will see that this property persists for more general systems. The fact that the emergent post-selected paths do not depend on quantum features of the system  suggests that they can also be defined  in setups where we do not know the underlying quantum theory, namely, quantum gravity. 

We differentiate Eq. (\ref{traj}) to obtain
\bey
\dot{\xi}^a = \{\xi^a, h\} + V^b L^{a}_b(t), \label{traj1b}
\eey
where 
\bey
L^a_b(t) = \frac{\partial f^a_t}{\partial  \xi^b}
\eey
is a time-dependent tensor on $\Gamma$ that describes the dependence of the solutions to the classical equations of motion on  the initial conditions. We will refer to $L^a_B$ as the Lyapunov tensor of the classical system. 
The effect of post-selection is contained in the constant vector
\bey
V^b = \frac{f^b_{-T}(\xi_f) - \xi^b_i}{T}. \label{postv}
\eey
Eq. (\ref{traj1b}) describes a combination of purely Hamiltonian evolution, as described by the term with the Poisson bracket, and evolution due to the post-selection conditions as given by $V_b L^{a}_b(t)$. This structure manifests a crucial property of history theories, that they admit two distinct notions of time translation, one generated by the Hamiltonian, and one corresponding to the histories'  intrinsic time ordering \cite{Sav99, Sav10}.

The only assumption involved in the derivation  of Eq. (\ref{traj}) is that the quantum system has a well-defined classical Hamiltonian limit,
as given by Eq.   (\ref{approxdet}). This is possible if the cells $V_{\alpha}$ of the partition remain regular under time evolution, i.e., that their boundaries do not develop structure at scales smaller than $\ell$. Systems with ergodic/mixing dynamics do develop such a structure, and this places a constraint on the total duration $T$, for which Eq. (\ref{traj}) applies.

 Eq. (\ref{traj}) does not conserve energy, even when $\xi_i$ and $\xi_f$ lie at the same energy surface. 
  As an example, consider a free particle of mass $m$ in one dimension, with $\xi_i = (x_i, p_i)$ and $\xi_f = (x_f, p_f)$. The post-selected trajectory is
 \bey
 x(t) = x_i + \left(\frac{x_f-x_i}{T}+ \frac{p_i - p_f}{m}\right)t  + \frac{p_f - p_i}{mT} t^2, \;\;\; 
 p(t) = p_i + \frac{p_f - p_i}{T} t \label{postpart}
 \eey  
If $p_f \neq p_i$, the particle undergoes constant acceleration $a = (p_f - p_i)/(mT)$.  
We also note that the velocity 
\bey
\dot{x}(t) = (x_f - x_i)/T + (p_i - p_f)/m + 2 \frac{p_f - p_i}{mT}t  \nonumber
\eey
 differs from $p(t)/m$. This is not paradoxical. Momenta and velocities correspond to different quantum observables, that are measured in different type of experiments; their difference is strongly highlighted in set-ups that involve post-selection. 
 
 Note that the post-selection vector (\ref{postv}) is $V = T^{-1}((L- p_f T/m), (p_f - p_i))$. If we take $L$ and $p_f$ to be functions on $T$, then $V$ remains finite at $T \rightarrow \infty$ if $L- p_f T/m$ and $p_f - p_i$ grow with $T$ as $T \rightarrow \infty$. Hence, final conditions can be meaningfully imposed also in the asymptotic future.
 
  Other examples of post-selected paths are presented in the Appendix B.

\subsection{Post-selection in a general state space}

We derived Eq. (\ref{traj}) for canonical coherent states, which describe a linear state space. There is no obstacle in generalizing this result to arbitrary state spaces. The only requirement is that the standard quantum evolution has a well defined classical Hamiltonian limit, in the sense that Eq. (\ref{approxdet}) holds. In this case, the analogue of Eq. (\ref{traj}) is
\bey
\xi(t) = f_t\left(\gamma_{\xi_i, \xi'_f}(t)\right), \label{traj2}
\eey 
where $\gamma_t(\xi_i, \xi'_f)$ is the geodesic connecting $\xi_i$ and $\xi'_f = f_{-T}(\xi_f)$, with respect to the metric (\ref{metric}) that is induced by the coherent states on the classical state space.

To obtain Eq. (\ref{traj2}), we use the expansion
\bey
\langle \xi + \delta \xi|\xi\rangle = \exp\left[ i \theta_a(\xi)\delta\xi^{\alpha} - \frac{1}{2} \gamma_{ab}(\xi) \delta \xi^a \delta \xi^b\right]. \label{xdx}
\eey
for the inner product of two coherent states with neighbouring arguments. In Eq. (\ref{xdx}), $\theta_a = -i \langle \xi|\partial_a|\xi\rangle$ is the Liouville one-form for the Hamiltonian system.   Using Eq. (\ref{xdx}), the analogue of Eq, (\ref{AVN}) for sufficiently large $N$ is 
\bey
{\cal A}(V_1,t_1; V_2,t_2; \ldots; V_{N-1}, t_{N-1}) 
= \int_{f_{-t_1}(V_1)} d\xi_1 \int_{f_{-t_2}(V_2)} d\xi_2 \ldots  \int_{f_{-t_{N-1}}(V_{N-1})} d\xi_{N-1}  
\nonumber \\
\times \exp\left[i\sum_{n=0}^{N-1}\theta_a(\xi_n) \delta \xi^a_n- \frac{1}{2} \sum_{n=0}^{N-1} g_{ab}(\xi_n) \delta \xi^a_n \delta \xi^b_n \right]  \label{AVN2}
\eey
 The first term in the exponent converges to the integral $\int \theta d\xi$, i.e., the Liouville term of the action. The symplectic form on $\Gamma$ is given by $\omega = d \theta$.

  To find a limiting expression for the second term in Eq. (\ref{AVN2}), we assume a path parameter $s$, such that $\xi_n = \xi(s_n)$ and $s_n - s_{n-1} = \tau$, where $\tau$ is a constant. That is, $s$ is a path parameter chosen so that the successive instants of a history occur after intervals of equal `duration'. Then, the second term in the integral approximates $ - \frac{1}{2\tau} \int ds \, g_{ab}(\xi) \, \dot{\xi}^a \dot{\xi}^b $. It is minimized by the geodesic $\gamma_{\xi_i, \xi_f}$ that connects the initial and final points. 

The analysis proceeds as in Sec. 2.4. Maximum probability is achieved for histories $(f_{-t_1}(V_1), f_{-t_2}(V_2)$, $\ldots, f_{-t_{N-1}}(V_{N-1}))$ that contain the geodesic $\gamma_{\xi_i, \xi_f'}$, the probabilities for all other histories being suppressed by a factor of $e^{-\ell^2}$. Thus, we obtain Eq. (\ref{traj2}) for a deterministic trajectory under post-selection. 

The differential equation corresponding to Eq. (\ref{traj2}) is
\bey
\dot{\xi}^a = \{\xi^a, h\} + V^b(t) L^a_b(t), \label{traj3}
\eey
where $L^a_b(t)$ is the Lyapunov tensor and $V_b(t)$ is the tangent vector to the geodesic connecting $\xi_i$ and $\xi_f$. 

The post-selected path depends on the metric $g$, which is inherited from quantum theory via the selection of coherent states. However, the detailed form of the metric $g$ does not enter Eq. (\ref{traj2}), but only its geodesics.  Different metrics may have the same  geodesics. For example, the phase space metrics for a particle in one dimension induced by the coherent states are of the form
\bey
ds^2 = \frac{dx^2}{L^2} + \frac{dp^2}{P^2},
\eey 
where $L$ and $P$ are scales for length and momentum respectively. The straight lines on the $(x, p)$ plane are geodesics to all  metrics of this form. The space of geodesics modulo reparameterization does not depend on $L$ and $P$. The affine parameter does depend on $L$ and $P$, but in Eq. (\ref{traj}) the geodesics appear with respect to  physical time $t$, and not the affine parameter. 

If the set of coherent states is induced by a representation of a Lie group $G$, then   $\Gamma$ is a homogeneous Riemannian space. The slightly stronger assumption that $\Gamma$ is a symmetric space\footnote{  A symmetric space is a homogeneous Riemannian space with the additional property that for each $p \in \Gamma$, there  is an isometry $\phi$ with $\phi(p) = p$ that acts as $-1$ on the tangent space $T_p\Gamma$ of $p$. Equivalently, in a symmetric Riemannian space the Riemann tensor is covariantly constant. 
Most physically relevant homogeneous state spaces are also symmetric spaces. }  guarantees that  all geodesics are curves derived from the action of $G$ on $\Gamma$\cite{Arvanito}. Hence, to construct the post-selected path, we only need to know how  $G$ acts on $\Gamma$. But this action can also be implemented through canonical transformations. This means that  post-selected paths are defined solely in terms of the geometry of classical Hamiltonian mechanics.

\subsection{Restricted post-selection }
So far, we have considered the case that there are distinct final conditions for all degrees of freedom of the system. It may be the case that only some degrees of freedom are fixed at the final time. We will consider this case in reference to the case of canonical coherent states of Sec 2.3.

Let us split the degrees of freedom $\xi \in \C^{\nu}$ as $\xi = \eta \oplus \zeta$, where $\eta \in \C^{\mu}$ and $\zeta \in \C^{\nu - \mu}$. We therefore write the coherent states as $|\eta, \zeta\rangle$. We choose a final state 
\bey
\hat{\rho}_f = \int d\zeta_f |\eta_f, \zeta_f\rangle \langle \eta_f, \zeta_f|.  
\eey
The probabilities $\mbox{Prob}(V_1,t_1; V_2,t_2; \ldots; V_{N-1}, t_{N-1})$ are obtained by integrating the right-hand-side of Eq. (\ref{probdis}), with respect to $\zeta_f$. 
For ease of notation, we will denote a generic history $(V_1, V_2, \ldots, V_{N-1})$ by $\alpha$.  
Since $\xi_i$ and $\eta_f$, are held fixed,  for each value of $\zeta_f$, there is a unique history $\alpha(\zeta_f)$ that dominates in Eq. (\ref{probdis}); the probabilities associated to all other histories are suppressed. We will also denote by 
\bey
f^*_{t}\alpha(z_f):= (f_{-t_1}(V_1(\zeta_f)), f_{-t_2}(V_2(\zeta_f)), \ldots, f_{-t_{N-1}}(V_{N-1}(\zeta_f))). \nonumber
\eey 
 It follows that
\bey
\mbox{Prob}(\alpha) \simeq K \int d \zeta_f  \left|\int_{f^*_{t}\alpha(\zeta_f)} \prod_{n=1}^{N-1}d\xi_n \exp\left(\frac{i}{2} \sum_{n=0}^{N-1} \mbox{Im}({\bf \xi}_n^*\cdot {\bf \xi}_{n+1}) - \frac{1}{2} \sum_{n=0}^{N-1} |\xi_n - \xi_{n+1}|^2 \right)\right|^2.
\eey
We showed that for fixed $\xi_i$ and $\xi_f$, the exponent in the integral is peaked around the path (\ref{xin}), which contributes a term 
\bey
 i \mbox{Im}({\bf \xi}_f^{\prime *}\cdot {\bf \xi}_i) - \frac{1}{2N} |\xi'_f - \xi_i|^2 =  i \mbox{Im}({\bf \eta}_f^{\prime *}\cdot {\bf \eta}_i)  + i \mbox{Im}({\bf \zeta}_f^{\prime *}\cdot {\bf \zeta}_i) -\frac{1}{2N} \left( |\eta'_f - \eta_i|^2 + |\zeta'_f - \zeta_i|^2 \right) \nonumber
\eey
to the exponential.

When keeping $\eta_i, \zeta_i$, and $\eta_f$ fixed, and vary $\zeta_f$, both real and imaginary part in the exponent are peaked for $\zeta'_f = \zeta_i$, or equivalently $\zeta_f = \zeta(T)$, where $\zeta(T)$ is the $\zeta$ component of $f_T(\xi_i)$, that is, the $\zeta$ component of the time evolution of the initial conditions. For sufficiently large coarse-graining scale $\ell$, we obtain approximate determinism with path (\ref{traj}) for $\zeta_f = \zeta(T)$. We conclude that any degrees of freedom that are not fixed at the final time are determined by the solution of Hamilton's equations for the specified initial conditions.

\subsection{General coarse-grainings}
The expressions for the post-selected paths derived in the previous sections are applicable to Hamiltonian coarse-graining, i.e., to set-ups that involve state space measurements that distinguish all degrees of freedom of the system. This is certainly not the most-general case. In many-body systems, we often have access only to a few special degrees of freedom. For example, the special degrees of freedom may  collective ones,  characterized by time-scales so large that they effectively decouple from the rest.  


In some cases, the inaccessible degrees of freedom may act as an environment that modifies the effective evolution equations for the system. Then, we work within the theory of open quantum systems \cite{OQS}. 
If the environment carries no memory, the evolution equations can be described in terms of an one-parameter family of completely positive maps $K_t$ on the space of quantum states. The defining probabilities for an $N$-time post-selected history is
\bey
\mbox{Prob}(V_1,t_1; V_2,t_2; \ldots; V_{N-1}, t_{N-1}) = \hspace{5cm}
\nonumber \\
CTr\left[\rho_f K_{t_N-t_{N-1}}[\hat{P}_{V_{N-1}}K_{t_{N-1}-t_{N-2}}[... K_{t_2 - t_1}[\hat{P}_{V_1}K_{t_1-t_0}[\hat{\rho}_i]\hat{P}_{V_1}]  ...] \hat{P}_{V_{N-1}}]    \right], 
\eey
for some constant $C > 0$. The method we employed in this section does not straightforwardly work in this case, especially since the maps $K_t$ are completely arbitrary. 

For maps $K_t$ that describe dissipative dynamics, we expect that the classical limit corresponds to classical dissipative equations of motion, and that this limit is obtained by coarse-graining at scales $\ell$ much larger than the characteristic noise of the system \cite{Ana96}. For example, if the environment is a thermal bath at temperature $T$, $\ell$ must be much larger than the scale of thermal fluctuations on state space. We conjecture that for such systems, post-selected paths would satisfy an equation analogous to (\ref{traj3}), but with dissipative rather than Hamiltonian equations of motion. 


\section{The decoherent histories framework}
Our past analysis relied on the rule (\ref{prepost}) for quantum probabilities in terms of both pre- and post-selection. This formula requires that the conditions for post-selection are expressed at a specific final moment of time. In general, post-selection can take different forms, for example, we may select by requiring that a specific event occurs at {\em some} time $T$. This means that we must add the contributions of different final times $T$ to the post-selected probability. Certainly, this cannot be done by integrating over $T$ in Eq. (\ref{prepost}), because $T$ does not appear as a random variable in that equation, but as an external parameter. 
 
A  more general framework for post-selection is provided by the decoherent histories approach to quantum theory and its variants 
 \cite{Gri, Omn1, Omn2, GeHa1, hartlelo, Ish94}. The benefits of this approach are the following.  (i) It enables the treatment of the classical limit beyond approximate determinism, i.e., the derivation of stochastic classical equations from quantum mechanics \cite{GeHa2}. (ii) It can describe more general types of selection, including selection at intermediate moments of time, and selection at unspecified moments of time \cite{AnSav17}. (iii) It can be applied to a quantum cosmological setting, because it purports to describe closed systems with no reference to measurement at a fundamental level. In this paper, we will not explore directions (i) and (ii) in relation to post-selection. The main purpose of this section is to introduce decoherent histories as an appropriate framework for discussing post-selection in quantum cosmology. Hence, this short section contains almost only review material.

The basic object is a
    {\em history}, that is, a sequence of  properties of a physical
system at successive instants of time.  An $N$-time history
$\alpha$
correspond to a sequence $\hat{P}^{(1)}_{a_1}, \hat{P}_{a_2}^{(2)}, \ldots \hat{P}^{(N)}_{a_N}$ of projectors
each corresponding to an outcome $a_i$ in the measurement of an observable $\hat{A}^{(i)}$ at time $t_i$, where $i = 1, 2, \ldots, N$. Then, we construct the history operators \begin{eqnarray}
\hat{C}_{\alpha} =  \hat{P}^{(N)}_{a_N}(t_N) \ldots \hat{P}^{(2)}_{a_2}(t_2) \hat{P}^{(1)}_{a_1}(t_1), \label{classoper}
\end{eqnarray}
where $\hat{P}^{(i)}_{a_i}(t_i) : = e^{i\hat{H}(t_i-t_0)}\hat{P}^{(i)}_{a_i}e^{-i\hat{H}(t_i - t_0)}$ is the Heisenberg-picture evolution of $\hat{P}^{(i)}_{a_i}$. 
The probability associated to a history is
\begin{equation}
p(\alpha) = Tr \left( \hat{C}_{\alpha}\hat{\rho}_0  \hat{C}_{\alpha}^{\dagger} \right). \label{histprob2}
\end{equation}

The decoherent histories approach starts from the re-interpretation of histories as a sequence of properties of a physical system rather than as a sequence of measurement outcomes. The benefit   is that histories-as-properties  have a powerful logical structure. We can make propositions about histories and relate these propositions by logical operations like AND ($\wedge$), OR ($\vee$) , NOT ($\neg$), and so on. The set of history propositions forms a lattice, and it includes the trivially true proposition $I$ and the trivially false proposition $\emptyset$. 

The price  is that Eq. (\ref{histprob2}) does not satisfy the Kolmogorov additivity condition
\bey
p(\alpha \vee \beta ) = p(\alpha) + p(\beta) \label{eqkol}
\eey
 for any pair of disjoint histories $\alpha$ and $\beta$.  

There is a partial resolution: we can define probability measures when restricting to specific sets of histories.  To this end, we first define the {\em decoherence functional} $d$, a complex-valued function of pairs of histories, as \index{decoherence functional}
\begin{equation}
d(\alpha, \beta) = Tr \left( \hat{C}_{\alpha} \hat{\rho}_0 \hat{C}_{\beta}^{\dagger}\right). \label{decfun}
\end{equation}
The diagonal elements $d(\alpha, \alpha)$ of the decoherence functional coincide with the probabilities $p(\alpha)$ of Eq. (\ref{histprob2}). 

Let $\Omega$ be an exclusive and exhaustive set of histories, that is, a set of histories  $\alpha_i$ labeled by an index $i$, such that $\alpha_i \wedge \alpha_j = \emptyset$ for $i \neq j$, and $\vee_i \alpha_i = I$. 
 If all histories in $\Omega$ satisfy the decoherence condition
\begin{equation}
 d(\alpha_i, \alpha_j) = 0, \;\;\; \mbox{for} \;\;\; i \neq j, \label{decc}
\end{equation}
then, Eq. (\ref{eqkol}) is satisfied, and we can define a probability measure on $\Omega$. Then, $\Omega$ is called a {\em consistent set} or a {\em framework}. 
In fact, the weaker condition $\mbox{Re} d(\alpha, \beta) = 0, \,\, \mbox{if}\,\, \alpha \neq \beta$, suffices for the definition of a framework, but the condition (\ref{decc}) is more appropriate for discussions of the classical limit. The classical limit is a  set of histories that  can be described approximately by a stochastic process; in this case, we can relax 
\emph{} condition (\ref{decc}) to that of approximate decoherence, $|d(\alpha, \beta)| < \epsilon << 1$ for all $\alpha \neq \beta$.

The logical structure of histories enables us to consider more general type of histories than  sequences of projectors \cite{Ish94}. If we denote the space of histories by ${\cal V}$, the decoherence functional is a map $d: {\cal V} \times {\cal V} \rightarrow \C$ that satisfies the following conditions.

\bigskip

\noindent 1. Normalization: $d(I, I) = 1$.

\medskip

\noindent 2. Hermiticity: $d(\beta, \alpha) = d^*(\alpha, \beta)$.

\medskip

\noindent 3. Linearity: $d(\alpha_1   \vee   \alpha_2, \beta) = d(\alpha_1, \beta) + d(\alpha_2, \beta)$, if $\alpha_1$ and $\alpha_2$ are disjoint.

\bigskip

This set of axioms is also satisfied by functionals that are not of the form (\ref{decfun})---see Refs. \cite{IL95, ILS} for a general characterization of the decoherence functionals and Ref. \cite{Ana03} for the conditions that lead to decoherence functionals of the form (\ref{decfun}).   Decoherence functionals  that are conditioned at both initial and final times are well defined; for histories that correspond to sequences of projectors they take the form 
\bey
d(\alpha, \beta) = \frac{ \mbox{Tr} \left( \hat{C}_{\alpha} \hat{\rho}_i \hat{C}^{\dagger}_{\beta} \hat{\rho}'_f\right)}{Tr(\hat{\rho}'_f \hat{\rho}_i)}, \label{dab0f}
\eey
where $\hat{\rho}_f' = e^{i\hat{H}T} \hat{\rho}_i e^{-i\hat{H}T}$.

It is straightforward to show that the set of histories considered in Secs. 2.3 and 2.4 defines a framework. The decoherence functional is
\bey
d(\alpha, \beta) = \frac{\langle \xi_f'|\hat{C}_{\alpha} | \xi_i\rangle \langle \xi_i | \hat{C}_{\beta}^{\dagger}|\xi'_f\rangle }{|\langle \xi_f'|\xi_i\rangle|^2}.
\eey
As shown in Sec. 2, one matrix element $\langle \xi_f'|\hat{C}_{\alpha} | \xi_i\rangle$ dominates over all other, the one that corresponds to a history that contains the path (\ref{traj2}), all other suppressed by a factor of $e^{-\ell^2}$. Hence, the condition for approximate decoherence is satisfied with an error $\epsilon$ of the order of the accuracy of Eq. (\ref{approxdet}). The decoherence condition is a consequence of approximate determinism.

\section{Post-selected dynamics in quantum cosmology}

\subsection{The scope of quantum cosmology}
Quantum cosmology is  a research program that purports to treat the Universe as a quantum system---for introductions, see \cite{Wilt, Halliwell}. It has multiple motivations: to resolve the cosmological singularity, to understand the origins of inflationary initial conditions and fine-tuning, to seek the quantum origins of the second law of thermodynamics, to explain the emergence of a classical spacetime and of observers subject to the rules of classical physics, and to test the consistency of interpretations of quantum theory when extended to the cosmological setting.

Quantum cosmology proceeds under two assumptions. First, we  assume  that gravity can be quantized in such a way that standard general relativity emerges at an appropriate limit. Second, we assume that the  formalism of quantum theory can be suitably upgraded to deal with a genuinely closed system. In the Copenhagen interpretation, physical systems are assumed to be in contact with a measuring apparatus external to the system. Hence, initial  and final states   are defined in terms of operations by the agents that perform the measurements. 

When treating the Universe as a quantum system, we cannot assume any external apparatus or prior prepatation. An alternative interpretation is not only desirable but essential.
Among existing interpretations of quantum theory, the most appropriate for cosmology are 
 Everett-type interpretations \cite{Everet1, Everet2} (including the Many-Worlds interpretation\cite{MWI1, MWI2}) and decoherent histories. The former are incompatible with any final cosmological condition. They focus on the universal quantum state, which evolves unitarily, and it unfolds into branches. Each branch  corresponds to a different ``world".
Probabilities are not fundamental; the wave function  provides a complete description of the Universe and it evolves deterministically. A final condition makes as little sense in this framework  as  in classical Newtonian mechanics.

In contrast, the decoherent histories approach can easily incorporate final conditions. The reason is that it treats the quantum state   as a component of the decoherence functional, and not as a fundamental object of the theory. Only a small subset of the space ${\cal D}$ of decoherence functionals are of the form (\ref{decfun}) that corresponds to the Schr\"odinger evolution of an initial state.
 The set ${\cal D}$ contains decoherence functionals with final states---as in Eq. (\ref{dab0f})---, others with intermediate restrictions, and yet others with more exotic forms \cite{ILS}. In decoherent histories,  all possible elements of ${\cal D}$ are candidates for the definition of the fundamental cosmological probabilities. We should exclude only those that are incompatible with observations.

An important benefit of the decoherent histories interpretation is that it decouples the issue of finding the quasiclassical equations of motion
 from long-standing problems in quantum foundations, such as  the emergence of a classical Universe  and of classical observers. The former only requires that we construct probabilities for an appropriate set of histories that satisfies a decoherence condition, and then finding the histories that maximize them. As explained in Sec. 3, the post-selected histories of Sec. 2 satisfy this condition.

For those reasons, we believe that the decoherent histories interpretation is the most appropriate for discussing post-selection in quantum cosmology. Nonetheless, it is possible to work with other interpretations.   In principle, we can   work with Bohmian mechanics, which admits a formulation with both initial and final states \cite{suther}.  Quantum interpretations that postulate some form of retrocausality \cite{retrorev} can also incorporate cosmological final conditions. In general, any approach to quantum theory that accepts probabilities at the fundamental level---e.g., stochastic hidden variable theories, dynamical collapse models---can  support final states in quantum cosmology, but the postulate of such states may conflict with the theory's avowed motivations and aims.

\subsection{The state space of quantum cosmology}
We will describe quantum cosmology using the Hamiltonian formulation of General Relativity (GR)---see, for example, \cite{MTW}. GR defines a parameterized system, that is, a constrained system with a Hamiltonian that vanishes due to first-class constraints. 

The classical state space $\Gamma$ of a cosmological model consists of the spatial
three-metric $h_{ij}$ on a three-surface $\Sigma$ and its conjugate momentum $\pi^{ij}$, as well as matter fields $\phi_r$, and their conjugate momenta $p^r$. We will describe the points of $\Gamma$ by the collective coordinate $\xi^a$.
 $\Gamma$ is equipped with the standard Poisson bracket  $\{,\}$. In cosmological models, one often restricts to field variables with a specific symmetry, for example, homogeneity and isotropy. In this case, the state space $\Gamma$ is finite dimensional. While we will use this approximation in this paper, we must point out that the effective evolution equations with post-selection derived in Sec. 2 do not depend on the dimensionality of $\Gamma$.

 The state space of GR is subject to constraints, which we denote collectively by $C_{\lambda}$, for some index $\lambda$. The constraints are first-class, i.e., they satisfy 
\bey
\{C_{\lambda}, C_{\lambda'}\}= f^{\mu}_{\lambda \lambda'} C_{\mu}, \label{fcc}
\eey
where $f^{\mu}_{\lambda \lambda'}$ are functions on $\Gamma$.

The submanifold $C$ of $\Gamma$, defined by the vanishing of the constraints is called the {\em constraint surface}. 
By virtue of Eq. (\ref{fcc}), the canonical transformations generated by the constraints preserve the constraint surface $C$. The action of these canonical transformations partitions  $C$ into 
mutually exclusive submanifolds, the constraint orbits. 
The true degrees of freedom of the system are contained in the set of all constraint orbits, the {\em reduced state space} 
 $\Gamma_{red}$. This is a symplectic manifold, with a symplectic form and Poisson bracket inherited from those of $\Gamma$.The Hamiltonian on $\Gamma_{red}$ vanishes due to the constraints. Hence, the solutions to the classical equations of motion on $\Gamma_{red}$ are simply $\xi(t) = \xi(0)$,  for any time parameter $t$.
Each point of $\Gamma_{red}$ corresponds to a solution of Einstein's equations, modulo diffeomorphisms.
  For the relations between paths on $\Gamma_{red}$ and solutions to the equations of motion, see the histories description of Hamiltonian General Relativity in \cite{Sav03a, Sav03b} 

When quantizing the systems, the true degrees of freedom are described in terms of a Hilbert space ${\cal H}$. The exact procedure of constructing ${\cal H}$ depends on the quantization program. A meaningful quantization scheme will map a set $Ob_C$ of classical observables  into a set $Ob_Q$ of self-adjoint operators, while preserving some algebraic structure of the classical state space. Both sets $Ob_C$ and $Ob_Q$ must be sufficiently large that they can ``generate" all observables of the classical state space and the quantum Hilbert space, respectively. 

A key point of our analysis of post-selected paths in Sec. 2 is that they can be expressed solely in terms of the geometry of the classical state space, which in this case is the reduced state space $\Gamma_{red}$. The details of the quantization scheme do not matter. Nonetheless, different quantization schemes identify different sets of coherent states, resulting to different  geometries on $\Gamma_{red}$.  This is to be expected, because different quantization schemes employ different sets $Ob_C$ and $Ob_Q$, resulting to different kinematics in the quantum-to-classical correspondence.

The constraints of GR include the generator of spatial diffeomorphisms. This means that the true degrees of freedom must be invariant under spatial diffeomorphisms. Typically these observables are non-local in space. The procedure through which one can recover local physics is a major conceptual problem of quantum gravity, usually subsumed under the more general label of the ``problem of time" \cite{IshamPT, Kuchar, Anderson}---see also the discussion in Sec. 4.5. Supposing that the quantum gravity theory solves this problem, the classical limit of a theory with  post-selection in all degrees of freedom would differ from GR even at the local scales of galaxies, or even the solar system. To avoid this, we must assume that any final conditions refer to a small set of global degrees of freedom $Ob_{ps} \subset Ob_C$ that have only cosmological significance. Then, the formalism of Sec. 2.6 applies. The  degrees of freedom on $Ob_{ps}$ evolve via the post-selected classical equations, while the remaining degrees of freedom follow 
Hamilton's equations of motion (that is, Einstein's equations). Thus, there is no conflict between post-selected dynamics in the cosmological scale and the validity of GR at the galactic and smaller scales.

Note that the restriction to a small number of degrees of freedom is not specific to our approach, but it reflects a defining  puzzling feature of modern cosmology. The Universe apparently starts with a homogeneous and isotropic metric, while initial conditions taken randomly could be any in the infinite dimensional manifold $\Gamma_{red}$ of GR. The cosmological initial conditions apparently ``activate" only a small number of available gravitational degrees of freedom, so this would have to be reflected also in the final conditions.

\subsection{Implementing post-selection in cosmology}

We assume a spacetime without boundaries, in which case the Hamiltonian on the Hilbert space ${\cal H}$ vanishes. Then, $f_t$ is the identity map: $f_t(\xi) = \xi$. Eq. (\ref{traj2}) yields the classical path 
\bey
\xi(\cdot) = \gamma_{\xi_i, \xi_f}(\cdot), \label{traj2b}
\eey 
where $\gamma_t(\xi_i, \xi_f)$ is the geodesic connecting $\xi_i$ and $\xi_f$, with respect to the metric on $\Gamma_{red}$  that is induced by the coherent states.

Eq. (\ref{traj2b}) applies for post-selection in any parameterized system. We will specialize to quantum cosmology of the 
Friedmann-Robertson-Walker (FRW) type, that is, we restrict to homogeneous and isotropic configurations for gravity and matter fields. We apply post-selection only to cosmological degrees of freedom, and ignore the rest. This means that we consider only observables at 
length scales larger than the cosmological  coarse-graining scale $L_{cg}$, which is of the order of $100Mpc$.

 We take the FRW metric
\bey
ds^2 = - N^2(\tau)d\tau^2 + a^2(t) d\sigma_3^2, \label{frw}
\eey
where $d \sigma_3^2$ is a  spatial three-metric of constant curvature,  $N$ is the lapse function, and $a$ the scale factor. The state space consists  of  $a$, its conjugate momentum 
\bey
\pi = - \frac{12 a \dot{a}}{N},
\eey
 together with matter degrees of freedom. We express the latter in terms of
 configuration coordinates $q_i$ and their conjugate momenta $p_i$. 

There is a single constraint, the Hamiltonian constraint, which takes the form
\bey
C = -\frac{\pi^2}{24 a} - 6 \kappa a + \rho a^3 = 0,
\eey
where $\kappa \in \{-1, 0, 1\}$, depending on the topology of the three-surfaces. 

Consider the case of an ideal fluid that is commonly employed in cosmology. 
The energy density $\rho$ is a function of the number densities $n_s$ for particles of different species labeled by $s = 1, 2, \ldots, N$. For ease of notation, we will represent the number densities by the vector ${\bf n} = (n_1, n_2, \ldots, n_N)$. For a perfect fluid, with conserved number of particles in each species, ${\bf n} = \bfnu
\sqrt{h^0}/\sqrt{h} = \bfnu/a^3$, where $\bfnu$ is a constant vector, and $h^0$ is the determinant of the metric $d\sigma_3^2$.

In the Hamiltonian description of gravitating perfect fluids, $\bfnu$ is treated as a fixed parameter. Here, we follow the formalism and conventions of Ref. \cite{Kij}.
 When we assume that the perfect fluid  emerges from the more fundamental degrees of freedom $q_a, p_a$,  $\bfnu$ is a dynamically conserved {\em variable}, that is, it depends on   $q_a, p_a$, and  $a$, and it  satisfies
  $\{C, \nu_a\} = 0$ on the constraint surface. For fixed $\bfnu$, the reduced state space contains a single point: there is a unique state space trajectory for each vector $\bfnu$, which corresponds to a unique FRW solution.
   This means that the components $\nu_s$ are coordinates on the reduced state space $\Gamma_{red}$, and that the space $Z$ of all vectors $\bfnu$ is a submanifold of $\Gamma_{red}$.
 
Suppose that the  initial and final conditions  correspond, respectively, to $\bfnu_0$ and $\bfnu_f$ on $\Gamma_{red}$. We denote the corresponding quasi-classical path by $\bfnu(\tau)$, where  $\tau$ is the time parameter in the metric (\ref{frw}). 
The simplest choice is to take $\tau = t$, the proper time of isotropic observers in the FRW spacetime. In this case $N = 1$, and 
the constraint equation becomes
\bey
     \dot{a}^2 +\kappa  = \frac{1}{6} a^2 \rho\left[\bfnu(t)/a^3\right]. \label{frw1}
\eey
The only difference from the standard FRW equation is the time-dependence of $\bfnu$. By differentiation, we obtain
\bey
\frac{\ddot{a}}{a} = - \frac{1}{12}(\rho + 3 P) + \frac{1}{12 a} \frac{\partial \rho}{\partial {\bf n}} \cdot \dot{\bfnu}, \label{frw2}
\eey
where we used the thermodynamic identity for the pressure $P =  {\bf n}\cdot \frac{\partial \rho}{\partial {\bf n}} - \rho$.

Suppose that $\rho$ is of the form $\sum_s \rho_s(n_s)$, i.e., the different species contribute additively to the total energy density. Then, we can define the partial pressure $P_s = n_s(\partial \rho_s/\partial n_s) - \rho_s$ for the $s$-species, so that $P = \sum_s P_s$.  Eq. (\ref{frw2}) becomes
\bey
\frac{\ddot{a}}{a} = - \frac{1}{12}(\rho + 3 P) + \sum_s c_s (\rho_s + P_s), \label{frw3}
\eey
where 
\bey
c_s = \frac{d \ln \nu_s}{d \ln a},
\eey
is the logarithmic derivative of $\nu_s$.

Eq. (\ref{frw3}) can also be written as
\bey
\frac{\ddot{a}}{a} = \frac{1}{12} \sum_s \left[(c_s - 1) \rho_s + (c_s - 3)P_s \right].
\eey
The contribution of any species that satisfies the dominant energy condition ($P_s \leq \rho_s$) is positive if $c_s > 2$. For dust ($P_s = 0$), we have acceleration if $c_s > 1$;  for radiation ($P_s = \frac{1}{3}\rho_s$), we have acceleration if  $c_s > 3/2$.

\subsection{A model calculation}

The determination of $\bfnu$ as a function of $a$ requires an input of quantum gravity theory, in the form of a set of coherent states in the physical Hilbert space. We lack a quantum gravity theory, but we can see how this works by considering a simple model. 

Consider a FRW spacetime with dust, in which case $\rho = c n$. As long as the scale factor is an increasing function of time $t$, we can express the lapse $N$ as a function of $a$. The form of the lapse function depends on the choice of the time variable  $\tau$.  If we identify $\tau$ with the proper time $t$ of isotropic observers, then  $N(a) = 1$. 
If we identify $\tau$ with the conformal time $\bar{t} = \int^t dt'/a(t') $, then $N(a) = a$.  In what follows, we will focus on lapse functions given by a power-law $N(a) = a^s$. 

We define
\bey
x = \int^a \frac{\sqrt{a'} da'}{N(a')},
\eey
and we denote by $a(x)$ the inverse function, expressing $a$ as a function of $x$. For $N(a) = a^s$, $a(x) = [(\frac{3}{2}-s) x]^{\frac{1}{\frac{3}{2} - s}}$.

The FRW equations become
\bey
\dot{x}^2 + \kappa a(x) = \frac{c\nu}{6}.
\eey
The parameter $\nu$ is therefore proportional to the energy of a particle in a line moving  in a potential  $V(x) = 2 \kappa a(x)$. The reduced state space that includes different values of $\nu$ is spanned by the canonical pair of variables $x$ and $p = \dot{x}$. 

We will specialize to the case of $\kappa = 0$, the cases of $\kappa = \pm 1$ are straightforward generalizations. The canonical transformation corresponding to the equations of motion is that of a free particle, $f_{\tau}(x, p) = (x + p \tau, p)$. If we assume that the associated quantum theory supports the canonical coherent states for the variables $x$ and $p$, then the state space metric is of the form
\bey
ds^2 = \beta^2 dx^2+ \beta^{-2} dp^2, \label{metfl}
\eey
for some constant $\beta$. The 
post-selected path is given by Eq. (\ref{postpart}) for $m = 1$. Hence, $x(\tau)$ grows with $\tau^2$ at sufficiently late times. This implies that 
\bey
t \sim \tau^{\frac{\frac{3}{2} +s}{\frac{3}{2} -s}}, \nonumber
\eey
and that $a(t) \sim t^{\frac{2}{\frac{3}{2}+s}}$. We have cosmic acceleration if $\frac{d^2a}{dt^2} >0$, i.e., for $s < \frac{1}{2}$. 

The canonical coherent states may not be the most appropriate for the quantization of this system, because they do not take into account the fact that $x$ takes only positive values, that is, the configuration space is the half-line and not  the full real line. The 
reduced state space is $ T^*R^+ =  \R^+\otimes \R$, and it consists of  canonical pairs
$(x, p)$ with  $x > 0$. The canonical commutation relations $[\hat{x}, \hat{p}]= i \hat{I}$ are not well defined for a particle in the half-line; by the Stone-von Neumann theorem, the spectrum of $\hat{x}$ must be the full real line---see, for example, Ref. \cite{Ishb} 
 The natural quantization algebra for a particle in the half-line is the    affine algebra 
\bey
[\hat{x}, \hat{\pi}] = i\hat{x}.
\eey
The representation of this algebra on the Hilbert space generates a set of affine coherent states \cite{Klaub}. The associated state space metric is
\bey
ds^2 =  \frac{dx^2}{x^2} + x^2 d p^2 = dq^2 + e^{q} dp^2, \label{afme}
\eey
where we wrote $x = \beta e^{q/2}$ for some constant $\beta$. We  derive the geodesic equations
\bey
\ddot{q} =  \frac{1}{2}e^{q} \dot{p}^2,\;\;\; \frac{d}{d\tau} \left(\dot{p}e^{q}\right) = 0.
\eey
Then $\dot{p}e^{q} = c$, and the geodesic equation for $q$ becomes
\bey
\frac{1}{2} \dot{q}^2 + V(q) = \epsilon \label{parpot}
\eey
where $V(q) = \frac{c^2}{2} e^{-q}$, and  $\epsilon$ is an integration constant. It is straightforward to evaluate the geodesics,
\bey
x(\tau) = x_i \cosh\left(b\tau\right), \;\;\; p(\tau) = p_i + \frac{p_f - p_i}{\tanh(bT)}\tanh(b\tau ),
\eey
where $b = T^{-1}\cosh^{-1}(x_f/x_i)$. 

Hence, the post-selected paths are 
\bey
x(\tau) &=& x_i \cosh\left(b'\tau\right) + \left[p_i + \frac{p_f - p_i}{\tanh(b'T)}\tanh(b'\tau)\right] \tau , \\
p(\tau) &=& p_i + \frac{p_f - p_i}{\tanh(b'T)}\tanh(b'\tau),
\eey
where $b' = T^{-1}\cosh^{-1}\left(\frac{x_f - p_fT}{x_i}\right)$.

At sufficiently large $t$, $x(t) \sim e^{\frac{1}{2}b't}$ and $p(t)$ is constant. In this regime,
\bey
a(t) \sim \left\{ \begin{array}{cc} e^{\frac{1}{3}b' t},& s = 0 \\ t^{s^{-1}}, & s > 0, \end{array} \right.,    \label{ataf}
\eey
that is, we have cosmic acceleration for $s < 1$. Note that for $s = 0$, the spacetime is asymptotically de Sitter, with a Hubble factor   $\frac{1}{3}b'$.

Yet another quantization is obtained by considering the Poisson algebra generated by the functions $K_0 = x$, $K_1 := x \cos p $, and $K_2 = x \sin p$,
\bey
\{K_0, K_1\} = K_2, \;\; \{K_2, K_0\} = K_1, \;\; \{K_1, K_2\} = -K_3.
\eey
This is isomorphic to the algebra ${\it sl}(2, \R)$; the associated Casimir is $C_2 = K_0^2- K_1^2 - K_2^2 = 0$. A quantization based upon unitary representations of the group $SL(2, \R)$ mimics loop quantum gravity quantization of mini-superspace models \cite{Bojo}. In particular, it yields discrete spectrum for the scale factor of the Universe. The $SL(2, \R)$-invariant metric induced by the coherent states of $SL(2, \R)$ \cite{Perelomov} is 
\bey
ds^2 =   dq^2 +  \beta^{-2} \sinh^2 q \, dp^2, \label{sl2rm}
\eey
where $x = \beta \cosh q$, for some constant $\beta$. It is straightforward to show that the geodesic equations lead again to Eq. (\ref{parpot}), with a potential $V(q) = \frac{c^2}{\sinh^2q}$, for some constant $c$. The solutions are 
\bey
x(\tau) = x_i \cosh\left(b\tau\right), \;\;\; p(\tau) = p_i + \frac{p_f - p_i }{\tan^{-1}\left(\kappa\tanh(bT)\right)}\tan^{-1}\left(\kappa\tanh(b\tau)\right),
\eey
where  $b = T^{-1}\cosh^{-1}(x_f/x_i)$, and $\kappa = \sqrt{\beta^2x_i^2 - 1}$. 

The post-selected paths are
\bey
x(\tau) &=& x_i \cosh\left(b'\tau\right) + \left[p_i + \frac{p_f - p_i }{\tan^{-1}\left(\kappa\tanh(b'T)\right)}\tan^{-1}\left(\kappa\tanh(b'\tau)\right)\right]\tau\\
p(\tau) &=& p_i + \frac{p_f - p_i }{\tan^{-1}\left(\kappa\tanh(b'T)\right)}\tan^{-1}\left(\kappa\tanh(b'\tau)\right).
\eey
Post-selected paths at large times behave $x(t) \sim e^{\frac{1}{2}b't}$ and $p(t)$ is constant. Then,  Eq. (\ref{ataf}) applies. 

\subsection{The input from quantum gravity}

The models of Sec. 4.4 provide specific predictions about cosmological evolution in presence of post-selection with minimal information from a quantum theory of gravity. The predictions are not unique, but this reflects solely our ignorance about the correct theory of quantum gravity. 

In general, there is an ambiguity due to the choice of the time variable. The choice of time in GR is equivalent to fixing a  gauge. In the classical theory, determinism and diffeomorphism invariance guarantee that physical properties are independent of the gauge conditions that define the time parameter. No such guarantee exists in quantum theory; in fact,  quantum gravity theories  with  different time parameters are expected to  have unitarily inequivalent dynamics \cite{Haj1, Haj2, Malk}. This is a crucial  aspect of the problem of time in quantum gravity. It is no surprise that properties of the deep quantum regimes, such as the post-selected paths, are strongly dependent on the time parameter. 

While the problem of time remains a challenge for all quantum gravity programs, it is not particularly acute for the models presented here, which do possess a preferred time variable. There are three reasons. First, in dust spacetimes (and more generally, perfect fluid spacetimes), there is a preferred choice of time coordinate, corresponding to the proper times of the fluid's flow lines. Second, FRW spacetimes, being isotropic, also have a preferred time direction, defined by the proper times of isotropic observers. Third, the post-selected paths are defined with respect to a sequence of measurements, and the natural time parameter is one with respect to which the sequence of measurements is homogeneously distributed. If we are to compare the predictions with observational data, again, the proper time of isotropic observers is indicated.

Hence, the natural choice of time corresponds to $N = 1$, or, equivalently, $s = 0$. Then, post-selected paths behave as $a(t) \sim t^{4/3}$ for canonical coherent states and as $a(t) \sim e^{\frac{b'}{3} t}$ for affine coherent states and $SL(2, \R)$ coherent states. The different predictions correspond to different quantization schemes. Canonical coherent states follow from a  quantization based on a unitary representation of the canonical commutation relations. In the full state space of GR, these read   

\bey
[\hat{h}_{ij}({\bf x}), \hat{\pi}^{kl}({\bf x}')] = i \left(\delta_i{}^k \delta_j{}^l + \delta_i{}^l \delta_j{}^k\right) \delta^3({\bf x}, {\bf x}'). \label{ccr}
\eey
where $h_{ij}$ is a  three metric $h_{ij}$ on a manifold $\Sigma$ and $\pi^{ij}$ its conjugate momentum.
This is the starting point for traditional canonical quantization of gravity, and also of perturbative quantization.

Affine coherent states correspond to quantization based on unitary representations of the affine algebra
\bey
[\hat{h}_{ij}({\bf x}), \hat{\pi}^{k}{}_l({\bf x}')] = i \left(\delta_i{}^k \hat{h}_{jl}({\bf x}) + \delta_j{}^k \hat{h}_{il}({\bf x})\right)\delta^3({\bf x}, {\bf x}'). \label{aff}
\eey
This algebra represents more accurately the topological and global structure of the classical state space of GR \cite{IsKa}, and it defines an alternative starting point to canonical quantization \cite{Kla0, Pi, Kla3}. There are some differences from the analysis of Sec. 4.4.,  when reducing to homogeneous and isotropic cosmologies after quantization. These,   are described in the Appendix C.

The metric (\ref{sl2rm}) is adapted from a paradigmatic quantum cosmological model obtained from canonical quantization through the loop variables \footnote{Note a double meaning of the phrase ``canonical quantization". It can be used to refer to quantization via the canonical commutation relation (\ref{ccr}), or to quantization based on Hamiltonian methods, in contrast to path integral quantization. Here, we use the word in the latter sense. Hence, we view Eqs. (\ref{ccr}), (\ref{aff}) and the loop algebra as starting points of different approaches to canonical quantization. }\cite{Bojo}. Note that this algebra misrepresents the global structure of the classical state space, in the sense that it does not correspond to a transitive group action on the state space. 

Our analysis demonstrated that  different approaches to the quantization of gravity generate different post-selected paths. This is not a surprise: the difference between quantizations lies at the level of the operators that represent classical observables. Quantization via Eq. (\ref{ccr}) leads to a representation of the scale factor  by an operator whose  spectrum is the whole real line; quantization via Eq. (\ref{aff}) leads to an operator for the scale factor with spectrum the half-line; 
loop quantization leads to an operator for the scale factor with discrete spectrum. This means that the classical-quantum correspondence is different in different theories. Since post-selected paths are not part of the usual quasi-classical domain---where we can expect different approaches to converge---, but they arise from rare quantum processes, the dependence of the results on the quantization scheme is natural. Therefore, cosmological post-selection sharply distinguishes between competing quantum gravity programs.

One limitation of the models in Sec. 4.4 is that they treat matter as a perfect fluid. The perfect fluid description emerges at the classical limit, and that it is not fundamental.  A first-principles analysis should involve a full quantum description of matter via quantum field theory. This would necessitate going beyond the FRW approximation, 
 because the derivation of the hydrodynamic limit requires the full spacetime dependence of the field. The price of our simplified analysis is that the Hamiltonian description of fluids is disconnected from an underlying microscopic theory, and we do not ignore  effects of the quantum nature of matter on the post-selected paths.

\subsection{An interpolating parameterization } 

Since the post-selected paths are dependent on the underlying quantum gravity theory, it is useful to have a theory-independent parameterization of the paths $\bfnu(t)$ in Eq. (\ref{frw1}). This can be achieved by a simple interpolation between an initial value $\bfnu_i$ at scale factor $a_i$ and a final value $\bfnu_f$ at scale factor $a_f$. We assume that $a_i$ corresponds to  the very early Universe---at the earliest time that we can meaningful talk about a quasi-classical behavior of spacetime---, and that $a_f$ is at least as large as the present scale factor. For values of $a$ close to those of the present era, the scale difference is so enormous that we can take $a_i \simeq 0$. 

Suppressing the index $s$, we write a power-law interpolation as
\bey
\nu(a) = \nu_i + (\nu_f - \nu_i) (a/a_f)^{\lambda},   \label{interpolation}
\eey
where $\lambda$ is an exponent that may differ from species to species, and we take $v_f > v_i$. Note that in this interpolation, we use $a$ as a clock, i.e., we assume that it is a monotonous function of the time parameter $\tau$.

We readily calculate
\bey
c(a) = \frac{\lambda}{1 + \frac{\nu_i}{\nu_f - \nu_i}\left(\frac{a_f}{a}\right)^{\lambda}}.
\eey
The function $c$ increases from zero at $a = 0$ to its maximum value $\lambda(1 - \nu_i/\nu_f)$ at $a = a_f$. 
 
 In a dust cosmology,
 \bey
     \dot{a}^2 +\kappa  = \frac{c\nu_i}{6a } + \frac{c (\nu_f - \nu_i)}{6a_f^{\lambda}} a^{\lambda - 1}, \label{frw5}
\eey
where we used the time parameter $t$ of isotropic observers. 
For $\kappa = 0 , -1$, $\dot{a}$ never vanishes. It follows that $a$ is an increasing function of time, so the interpolation (\ref{interpolation}) is valid for all $a \leq a_f$. For $\kappa = 1$, the necessary condition for $\dot{a}$ to never vanish is $\lambda > 1$, and  
\bey
\frac{c\lambda \nu_i}{6 \alpha_f}\frac{(\nu_f/\nu_i-1)^{1/\lambda}}{(\lambda - 1)^{1- 1/\lambda}} > 1.
\eey
We readily observe that the last term in the right-hand side of Eq. (\ref{frw5}) is equivalent to   a cosmological fluid with equation of state $P = -\frac{\lambda}{3}\rho$, i.e., to a dark-energy fluid. A cosmological constant corresponds to  
 $\lambda  = 3$, and it has the same sign as $\nu_f - \nu_i$.
  The exponent $\lambda$ can be larger than $3$, so effective equations of state with $P/\rho < -1$---seemingly violating the dominant energy condition---are, in principle, possible. 
Note that if $\nu_i$ and $\nu_f$ are of the same order of magnitude, the matter term and the effective dark energy term are always of the same order of magnitude, so there is no cosmic coincidence problem.

If we treat $\nu_f - \nu_i$ a function of $a_f$ that grows with $a_f^{\lambda}$, the boundary condition remains non-trivial even for $a_f \rightarrow \infty$. Hence, the final condition can be thought of as being imposed at the spacetime's asymptotic future $\mathcal{I}^+$.

At large times, Eq. (\ref{frw5}) gives $a(t) \sim t^{\frac{2}{3-\lambda}}$ for $0 < \lambda < 3$, and $a(t) \sim e^{ht}$ for some constant $h > 0$ if $\lambda = 3$. Hence, the post-selected cosmology obtained from the canonical commutation relation has the same asymptotic behavior as Eq. (\ref{frw5}) while $\lambda = \frac{3}{2}$. Similarly, the cosmology obtained from the affine or the ${\it sl}(2, \R)$ algebras corresponds to $\lambda = 3$.

In a radiation-dominated cosmology, $\rho = c' n^{4/3}$ for some constant $c'> 0$. Then, Eq. (\ref{frw1}) yields
\bey
 \dot{a}^2 +\kappa  = \frac{c'\nu_i^{4/3}}{6a^2 } \left[1 + \left(\frac{\nu_f}{\nu_i} - 1\right)\left(\frac{a}{a_f}\right)^{\lambda} \right]^{4/3}, \label{rdc}
\eey
with no analogues in existing models for dark energy. 


The modification of dynamics due to post-selection is not physically equivalent with   a modified equation of state. Certainly, any arbitrary dependence of the energy density $\rho$ on $a$ in Eq. (\ref{frw1}) can be reproduced by a specific functional dependence of the pressure $P$ on $\rho$. (This function $P(\rho)$ may be physically problematic: as is well known, to account for cosmic acceleration, the modified equations of state must have negative pressure, and to violate the strong energy condition.) However, in thermodynamics an equation of state $P(\rho)$ enables us to derive the temperature $T$ as a function of $\rho$, thus, giving rise to a cosmic history of temperature $T(t)$.  The point is that the thermodynamics of matter does not change in presence of post-selection. For example, for radiation $T$ is always proportional to $\rho^{1/4}$.  
Hence, even if the evolution $a_{ps}(t)$ of the scale factor in a system with post-selection coincides with the evolution $a_{eos}(t)$ for a modified equation of state, the corresponding histories of cosmic temperature $T_{ps}(t)$ and $T_{eos}(t)$ can be very different.  

\section{Discussion}

Quantum post-selection enables us to probe rare quantum processes. In this paper, we showed that there is a hidden determinism among those rare processes. Given sufficient coarse-graining, we can derive quasi-classical evolution equations for generic quantum systems. These equations can be expressed solely in terms of classical physics, despite their quantum origin. 

In the cosmological context, we argued that if one 
\begin{enumerate}[(i)]
\item accepts that it is meaningful to describe the Universe as a quantum system, and 
\item is not committed to the Many-Worlds or other Everettian interpretations of quantum theory,
\end{enumerate}
there is no reason to restrict to only initial states  in  quantum cosmology. {\em A priori}, final states or even other conditions are equally plausible. Unlike classical physics, quantum physics needs not be conceptualized in terms of the evolution of initial data, and theories of quantum cosmology should consider all possibilities compatible with the mathematical structure of quantum theory. Using the quasi-classical equations for post-selected systems in FRW cosmology, we found that cosmic acceleration is possible  in absence of a cosmological constant, dark energy, or modified gravity.

We proceed to discuss the interpretation and implication of our results.

\bigskip

\noindent{\em Initial/final states as laws of nature.} The complete description of physical systems  requires the knowledge of its initial conditions and its evolution laws. Fundamental physical theories usually focus on the latter. Physical laws about initial conditions are only meaningful in the cosmological context. Such laws include 
 Boltzmann's explanation of the origin of irreversibility by a cosmological initial condition  and Ritz's cosmological explanation of time arrow of radiation \cite{Zeh}.

Quantum cosmology rekindled the interest  on theories about the initial condition of the Universe. Hartle and Hawking's no boundary proposal \cite{HH83} is the first example of a single physical law---summing over all spacetime geometries with no boundary in the past---that determines both the dynamics and the initial conditions of the Universe. This property is also shared by the causal set program to quantum gravity \cite{BLMS,Surya}.  

In our opinion,  final conditions in quantum cosmology should be viewed as a component of a single fundamental law that also determines initial conditions and dynamics. In the decoherent histories approach, this means that a physical law should determine the decoherence functional for the Universe. In this viewpoint, the post-selected probabilities are absolute rather than conditional. However, the formalism can also be used for conditional probabilities. In this case, post-selection is not a law of nature but the incorporation of information about the present state of the Universe, as in the anthropic principle.

The cosmological final conditions considered in this paper make no claim of fundamental nature. We focused on observable consequences of a final cosmological condition, and not on the grounding of such a condition on a more fundamental theory. While a fundamental description likely requires a quantum theory of gravity, we can still work with minispuperspace models that are expressed in terms of a finite number of degrees of freedom. In this context, our next aim is to define a quantum theory subject to initial and final cosmological conditions at the conformal boundaries of a cosmological spacetime, namely, at the initial singularity $\mathcal{I}^-$, and at the asymptotic future $\mathcal{I}^+$. Such a theory would be the basis of a first-principles analysis of cosmological perturbations in presence of post-selection.

\bigskip

\noindent {\em Causality and locality.} Any theory with post-selection must beware of  retrocausality, that is, the influence of future actions on past events. The final cosmological condition itself is not a problem. It is part of the probabilistic structure of the Universe, and as such it is fixed and unalterable by the actions of any agent. 

However, retrocausality may take other forms. Kent argued that some cosmologies with final conditions may allow superluminal signals \cite{Kent}. These superluminal signals are rather weak, as they cannot used to form closed causal loops. The situation considered by Kent assumes that the final condition can be taken for localized degrees of freedom, so it is not relevant to our model, in which  the final conditions are restricted to a small number of globally defined observables.

Certainly, it is important to undertake a detailed analysis of all possible forms of retro-causality in post-selected cosmologies. However, we believe that the theory will not exhibit strong causal pathologies if expressed in the language of histories. History-based theories incorporate the logical arrow of time in the very definition of the notion of a history, so, by construction, they cannot involve  closed causal loops.

\bigskip

\noindent {\em Teleology and empirical science.} A final condition in cosmology can be viewed as a teleological law of physics, that is, a law that asserts that the Universe will bring about a specific state of affairs. It is often asserted that teleology is outside the purview of science, because it is fundamentally untestable. Our results demonstrate that this is not always the case. Final conditions may lead to evolution equations that strongly differ from the expected classical equations of motion, and in the cosmological setting, they can account for cosmic acceleration. 

The usual explanations of cosmic acceleration postulate a change in cosmological dynamics rather than a final condition. They invoke 
either additional degrees of freedom  (dark energy) or modified gravity. These dynamical modifications must be present at scales smaller than the cosmological ones. 
  In contrast, the effect of final conditions is only seen at cosmological scales. At smaller scales, the universe is described by classical GR without a cosmological constant, and by matter that satisfies the strong energy condition. Hence, it is possible, in principle, to distinguish between predictions based on post-selection from a quantum state and the predictions based on a change in dynamics. 

Progress towards a quantum theory of gravity will likely provide us with more sophisticated candidates for final conditions, 
 and these could lead to cosmological predictions that would be sharply distinguishable from those of other cosmological models. Such predictions need not refer only to cosmic acceleration. For example, they may relate to the origins of thermodynamic irreversibility and its manifestation in structure formation, to particle creation effects that leave an observational imprint in the present epoch \cite{Dav13}, or to macroscopic quantum correlations that arise as a result of post-selection.




\begin{appendix}

\section{Simple examples of emergent quasi-classical trajectories via post-selection}
In this section, we  present some simple examples  of quasi-classical equations of motion for systems subject to both initial and final conditions.

\subsection{Classical Brownian motion} \label{CBM}
First, we will consider a classical probabilistic system, a particle
 undergoing Brownian motion in one dimension. This system is described by a probability density $\rho(x)$, that evolves in time according to the diffusion equation
\bey
\partial_t \rho  = \frac{D}{2} \partial_x^2 \rho. \label{diffusion}
\eey
We assume the initial condition that $x = 0$ at $t = 0$ and the final condition $x = L > 0$ for $t = T$. The probability density for position at time $t \in [0, T]$ is 
$p_t(x) = g_t(x)   g_{T - t}(L - x)/g_T(L)$, where
\bey
g_t(x) = \sqrt{\frac{D}{2\pi t}} e^{-\frac{D}{2t}x^2},
\eey
is the propagator associated to Eq. (\ref{diffusion}). 
It follows that 
\bey
p_t(x) = C \exp\left[ -\frac{D}{2t}x^2 - \frac{D}{2(T - t)} (L - x)^2\right]
\eey
where $C$ is a normalization constant. This probability density is peaked at its mean
\bey
\langle x(t)\rangle = \frac{L}{T}t.
\eey
For sufficiently large $D$, the width of the peak is so small that the evolution is approximately deterministic. The particle follows the trajectory that corresponds to  
a free particle with velocity $L/T$. Evolution without final conditioning in which $\langle x(t)\rangle = 0$ and $\langle x(t)^2\rangle \sim t$; the distance of the origin for a typical path increases with $\sqrt{t}$.

For other examples of stochastic systems with initial and final conditions, see \cite{Schul4}.



\subsection{Classical limit from a peak in the probability distribution} \label{peaks}

Consider a free particle of mass $m$,  with Hamiltonian $\hat{H} = \frac{1}{2m} \hat{p}^2$. We assume a Gaussian initial state $\psi_0$, centered around $x = 0$ and with zero mean momentum, 
\bey
\psi_0(x) = \left(2 \pi \sigma^2\right)^{-1/4} e^{-\frac{x^2}{4\sigma^2}}. 
\eey 
We consider Gaussian position sampling at time $t$, with associated positive operators $\hat{P}_x = \int dx_1 \chi_{x}(x_1) |x_1\rangle \langle x_1|$, where $\chi_x(x_1) = (2\pi \delta^2)^{-1/2} \exp[-(x-x_1)^2/(2\delta^2)]$. The joint probability density for a measurement at time $t$ with outcome $x$ and a measurement at time $T$ with outcome $L$ is
\bey
p_t(x; L, T) = \langle \psi_0|e^{i\hat{H}t}\sqrt{\hat{P}_x} e^{i\hat{H}(T-t)} |L\rangle\langle L| e^{-i\hat{H}(T-t)} \sqrt{\hat{P}_x} e^{-i\hat{H}t}|\psi_0\rangle.
\eey  
For fixed $L$ and $T$, we straightforwardly find 
\bey
p_t(x; L, T) = C \exp\left[-\frac{x^2}{2 \delta^2} \left(1-\frac{a_t}{a_t^2+b_t^2}\right) +  \frac{2m}{T-t} \frac{b_t}{a_t^2+b_t^2}Lx  \right], 
\eey
where 
\bey
a_t = 1 + \frac{\delta^2}{\sigma^2 \left(1 + \frac{t^2}{4m^2\sigma^4}\right)}, \hspace{1cm} 
b_t = \frac{\delta^2}{\sigma^2} \left(\frac{\frac{t}{2m\sigma^2}}{1 + \frac{t^2}{4m^2\sigma^4}} +\frac{2m\sigma^2}{T-t} \right),
\eey
and $C$ is an $x$-independent term.

 For fixed $L$, the probability density is peaked around
\bey
x(t) = L \frac{2m\delta^2}{T - t} \frac{b_t}{a_t^2+b_t^2-a_t} \label{fpqp}
\eey
with a spread
\bey
\delta x(t) = \frac{\delta}{\sqrt{1+\frac{a_t}{a_t^2+b_t^2}}} < \delta.
\eey
For $x(t) >> \delta$, the probability density is strongly peaked around $x(t)$, as given (\ref{fpqp}). In this sense, 
Eq. (\ref{fpqp}) defines a quasi-classical path of the quantum system.
In comparison, the corresponding quasi-classical path for the same initial state and no final conditions is $x(t) = 0$. Obviously, the trajectory (\ref{fpqp}) is not a solution to the classical equations of motion, and it does not conserve energy. The path (\ref{fpqp}) emerges because the probability of the final condition for the given initial state is very small.

\begin{figure}[!tbp]
  \centering
  \includegraphics[width=0.45\textwidth]{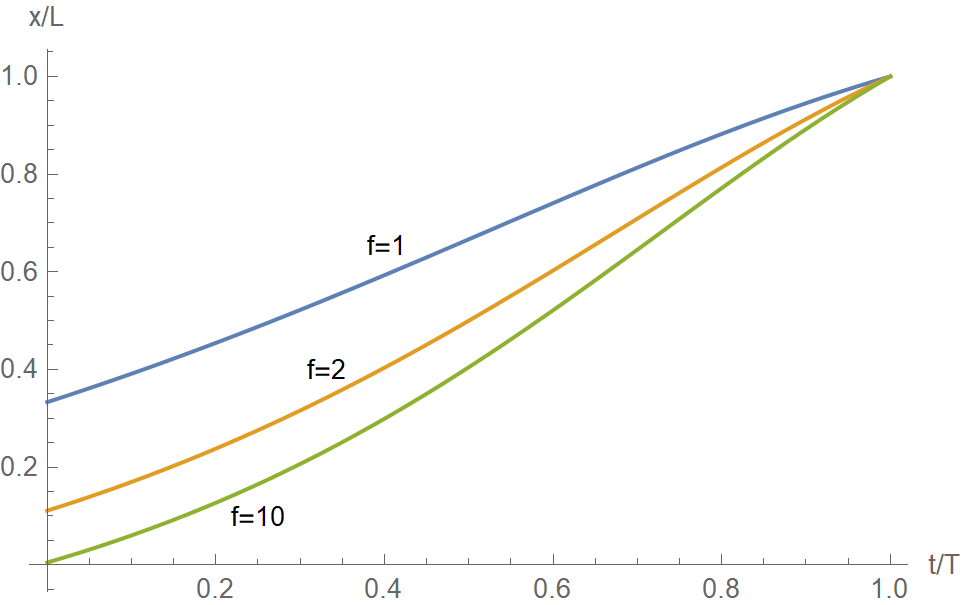}
  \caption{Plot of the quasi-classical path (\ref{fpqp}) for $\delta = \sigma$, and different values of the parameter $f$.}
  \label{fig1}
\end{figure}

The crucial parameter is $f = T/2m\sigma^2$, which quantifies the ratio of the duration of the evolution to the characteristic time parameter $2m\sigma^2$ of wave packet dispersion for a free particle.  In Fig. (\ref{fig1}), we plot the path (\ref{fpqp}) for different values of $f$. 
We find that for $f$ larger than about $5$, the path $x(t)$ is insensitive to $f$, while the dependence is stronger for smaller values of $f$.  

We also note that $x(0) = L \left[1+ f^2(1+\sigma^2/\delta^2)\right]^{-1}$.
For small $f$, $x(0)$ is far away from the locus of the initial state around $x = 0$. This   means that a very early measurement must record a very rare value of position in order to be compatible with the final condition.

 \section{Examples of post-selected paths}
 In this section, we give examples of some post-selected paths according to Eq. (\ref{traj}).
\subsection{Harmonic oscillator} 
\begin{figure}[!tbp]
  \centering
  \includegraphics[width=0.45\textwidth]{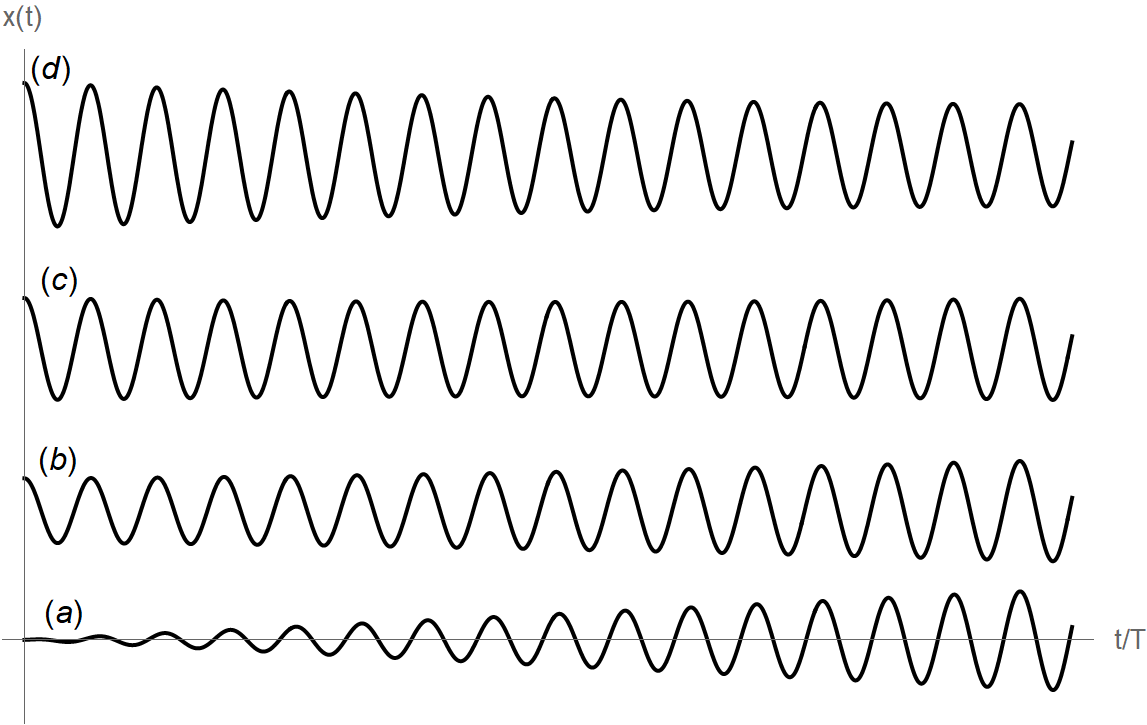}
  \caption{Plot of the position of the quasi-classical path (\ref{xtho}) for different values of the ratio of $E_i/E_f$: (a) $E_i/E_f = 0$, (b) $E_i/E_f = 0.5$, (c) $E-i/E_f = 1$, and (d) $E_i/E_f = 2$. In all plots, $\omega T = 100$, and $\phi = \pi/4$.}
  \label{plotxt}
\end{figure}

\begin{figure}[!tbp]
  \centering
  \includegraphics[width=0.45\textwidth]{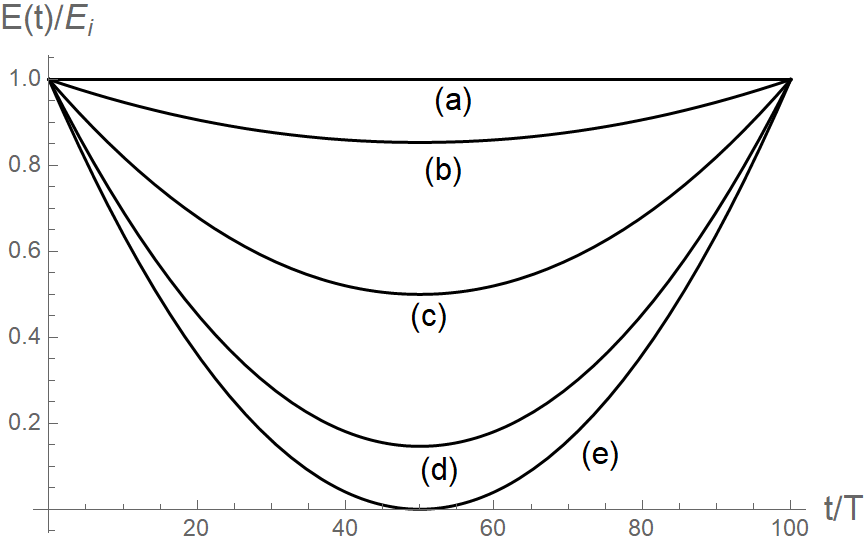}
  \caption{Plot of the energy of quasi-classical path (\ref{xtho}) for $E_i = E_f$ and for different values of $\phi$: (a) $\phi = 0$, (b) $\phi = \frac{\pi}{4}$, (c) $\phi = \frac{\pi}{2}$, (d) $\phi = \frac{3\pi}{2}$, and (3) $\phi = \pi$. In all plots, $\omega T = 100$.}
  \label{hoener}
\end{figure}

 Consider a harmonic oscillator in one dimension, in unites where the mass $m = 1$ and the frequency $\omega = 1$. The Hamiltonian is $h = \frac{1}{2}(p^2+x^2)$, and the solutions to the classical equations of motion are
 \bey
 f_t(x, p) = \left( \begin{array}{c} x \cos t + p \sin t\\-x \sin t + p \cos t\end{array}\right).
 \eey
Without loss of generality, we can take $\xi_i = (\sqrt{2E_i},0)$ and $\xi_f = (\sqrt{2E_f} \cos (\phi+T), \sqrt{2E_f}\sin(\phi + T))$, where $E_i$ and $E_f$ are the energy of the initial and final point, respectively. Then, the post-selected path is
\bey
x(t) = \sqrt{2E_i}\left(1 - \frac{t}{T}\right) \cos t + \sqrt{2E_f} \frac{t}{T} \cos(t - \phi)  \nonumber\\
p(t) = - \sqrt{2E_i} \left(1 - \frac{t}{T}\right) \sin t - \sqrt{2E_f} \frac{t}{T} \sin(t - \phi). \label{xtho}
\eey
A plot of $x(t)$ as a function of $t$ for different values of $E_i/E_f$ is given in Fig. \ref{plotxt}. In Fig. \ref{hoener}, we plot the total energy as a function of time for $E_i = E_f$. We see that energy is not conserved, and that is time evolution is sensitive to the phase of the final condition.
 
 \subsection{Crossing a classically forbidden region}
The fact that post-selected paths do not conserve energy, makes it possible to cross classically forbidden regions, that is regions, that cannot be crossed by a particle that starts from $\xi_i$ and evolves according to Hamilton's equations. The simplest example is provided by the inverse harmonic oscillator in one dimension. In units such that the Hamiltonian is $h = \frac{1}{2}(p^2 - x^2)$, the solution to the classical equations of motion are 
 \bey
 f_t(x, p) = \left( \begin{array}{c} x \cosh t + p \sinh t\\x \sinh t + p \cosh t\end{array}\right).
 \eey
 Consider an initial point $\xi_i = (-L, p)$ for $L > 0$. If $p < L$, a classical particle cannot cross to positive values of $x$. The region between $x = - \sqrt{L^2 - p^2}$ and $x = \sqrt{L^2 - p^2}$ is forbidden. To see how post-selected paths cross this region, take $\xi_f = f_T(L, p)$. The post-selected path is
 \bey
 x(t) = L\left(1 - \frac{2t}{T}\right) \cosh t + p \sinh t, \; \;\;\; p(t) = L\left(1 - \frac{2t}{T}\right) \sinh t + p \cosh t.
 \eey
 The associated energy is
 \bey
 E(t) = \frac{1}{2}\left[p^2 -L^2\left(1 - \frac{2t}{T}\right)^2\right],
 \eey
 i.e., it increases towards a maximum value $\frac{1}{2}p^2$ at $t = T/2$, and then decreases to its initial value at $x = L$. 
 
 \section{Metrics   in the state space of GR}
 
 The state space $\Gamma$ of pure GR without matter is the cotangent bundle of $\mbox{Riem}(\Sigma)$, where $\mbox{Riem}(\Sigma)$ is the space of Riemannian metrics $h_{ij}$ on a three-surface $\Sigma$. The Hamiltonian structure on $\Gamma$ is generated by the symplectic potential 
 \bey
 \Theta = \int d^3 x \pi^{ij}(x) \delta h_{ij}(x), \label{symppp}
 \eey
 where the tensor density $\pi^{ij}$ is the canonically conjugate variable to the metric $h_{ij}(x)$. This leads to fundamental Poisson brackets
 \bey
\{ h_{ij}({\bf x}),  \pi^{kl}({\bf x}')\} =   \left(\delta_i{}^k \delta_j{}^l + \delta_i{}^l \delta_j{}^k\right) \delta^3({\bf x}, {\bf x}'). \label{poisson}
\eey
Quantization via the canonical commutation relation (\ref{ccr}) requires that we extend the configuration space to the cotangent bundle of $T^{0,2}_S\Sigma$, where $T^{0,2}_S\Sigma$ is the space of $(0, 2)$ symmetric tensor fields on $\Sigma$. That is, we quantize by ignoring the metric positivity conditions on the tensors $h_{ij}$. In contrast, quantization via the affine canonical relations (\ref{aff}) preserves the configuration space of GR,  with positivity conditions. 

Either way, quantization induces---via coherent states---a metric on $\Gamma$ of the form
\bey
d s^2 = \int d^3 x \left( b^{-1}(x) M_{ijkl}(x)(\delta \pi^{ij}(x) \delta \pi^{kl}(x) + b(x) L^{ijkl}(x) \delta h_{ij}(x) \delta h_{kl}(x)\right), \label{metricGR}
\eey
where $b(x)$ is a reference density (a three-form) on $\Sigma$, and the tensors $M_{ijkl}$ and $L^{ijkl}$ are positive definite and they satisfy $L = M^{-1}$ as matrices. For quantization with the canonical algebra, $L^{ijkl}$ and $M_{ijkl}$ do not depend on the canonical variables. For quantization with the affine group,
$M_{ijkl}= \frac{1}{2}(h_{ik}h_{jl} +h_{il}h_{jk})$ and $L^{ijkl}= \frac{1}{2}(h^{ik}h^{jl} +h^{il}h^{jk})$.

  Suppose that we quantize the system of gravity interacting with a dust fluid, and specialize to the case of homogeneous and isotropic cosmology afterwards. Then, we have to pullback the symplectic potential (\ref{symppp}) and of the metric (\ref{metricGR}) to the state space of the FRW models. To this end, we 
we write   $h_{ij} = z \mathring{h}_{ij}$, where $\mathring{h}_{ij}$ is a reference homogeneous and isotropic metric on $\Sigma$. Then, the symplectic potential is pull-backed to $\xi d z$, if we express   $\pi^{ij}(x)$ as    $\frac{\xi}{\upsilon} b(x) \mathring{h}^{ij}(x)$, where   $\upsilon = 3 \int d^3x b(x)$.  

For quantization with the canonical algebra, we  obtain the metric
\bey
ds^2 = a_1 \delta z^2 + a_2 \delta \xi^2,
\eey 
with $a_1 = \upsilon^{-2} \int d^3x b(x) M_{ijkl}\mathring{h}^{ij}\mathring{h}^{kl}$ and $a_2 = \int d^3x b(x) L^{ijkl}\mathring{h}_{ij} \mathring{h}_{kl}$. 
The natural canonical coordinates are therefore $z$ and $\xi$, rather than $x$ and $p$ as in Sec 4.4. The geodesics are of the form 
\bey
z(t) = z_i + \frac{z_f - z_i}{T}\tau, \;\; \xi(t) = \xi_i + \frac{\xi_f - \xi_i}{T}\tau.
\eey
We straightforwardly find, that for the preferred time variable $t$, the post-selected paths at late times scale as $a(t) \sim t^{7/6}$.
 
 For quantization with the affine algebra, we obtain
 \bey
 ds^2 = \upsilon \frac{dz^2}{z^2} + z^{2} d\xi^2 = \upsilon dq^2 + \upsilon^{-1} e^q d \xi^2,
 \eey
 where $z - e^{z/2}$. This metric is similar to Eq. (\ref{afme}), but the specific coordinates on state space are different. This does not affect the asymptotic scaling of $a(t)$ as $e^{ht}$ for some constant $h$.
\end{appendix}

\end{document}